\documentclass[aps,showpacs,preprintnumbers,amsmath,amssymb]{revtex4}
\usepackage{dcolumn}
\usepackage[dvips]{epsfig}
\usepackage{graphicx}
\usepackage{amssymb}
\usepackage{color}
\usepackage{enumerate}
\usepackage{subfigure}
\usepackage{multirow}
\usepackage{amsmath}

\begin{document}
\baselineskip=0.8 cm
\title{\bf Thermal chaos of a charged dilaton-AdS black hole in the extended phase space}

\author{Chaoqun Dai$^{1}$,  Songbai Chen$^{1, 2 }$\footnote{Corresponding author: csb3752@hunnu.edu.cn},  Jiliang
Jing$^{1, 2 }$ \footnote{jljing@hunnu.edu.cn}}

\affiliation{$ ^1$Institute of Physics and Department of Physics, Key Laboratory of Low Dimensional Quantum Structures
and Quantum Control of Ministry of Education, Synergetic Innovation Center for Quantum Effects and Applications,
 Hunan
Normal University,   Changsha,  Hunan 410081,  People's Republic of
China \\
$ ^2$Center for Gravitation and Cosmology, College of Physical Science and Technology, Yangzhou University, Yangzhou 225009, China}

\begin{abstract}
\baselineskip=0.6 cm
\begin{center}
{\bf Abstract}
\end{center}

We have studied thermal chaotic behavior in the extended phase space for a charged dilaton-AdS black hole by Melnikov method and present the effect of dilaton parameter on the thermal chaos. Our result show that for the temporal perturbation the thermal chaos in the charged dilaton-AdS black hole occurs only if the perturbation amplitude is larger than certain a critical value, but for the spatially perturbation, the chaos always exists irrespective of perturbation amplitude. These behaviors are similar to those in other AdS black hole, which can be regarded as the common features of the static AdS black holes. Moreover, we also find that the critical temporal perturbation amplitude leading to chaos  increases with the dilaton parameter and decreases with the charge. This means that under the temporal perturbation the presence of dilaton parameter makes the onset of chaos more difficult, which differs from that of the charge parameter.

\end{abstract}

\pacs{ 04.70.Dy, 95.30.Sf, 97.60.Lf }
\maketitle

\newpage
\section{Introduction}

Chaos is a kind of very complicated and irregular motion, which appears only in the non-linear and non-integrable dynamical systems. It is highly sensitive to the initial conditions and presents the intrinsic random in systems \cite{Sprott,Ott,Brown1}. The chaotic phenomenon  can be
detected by many methods including the Poincar\'{e}
surfaces of section,  the Lyapunov characteristic exponents,  the fractal basin boundaries, the bifurcation diagram, the Melnikov method and so on. Although the chaotic motion is very complex, it
is very common in nature. In black hole physics, the chaotic orbits of particles  have been found to exist in multi-black hole spacetime \cite{Cornish,Hanan}, in the Manko-Novikov black hole spacetime \cite{Contopoulos,Contopoulos1,Contopoulos102,Contopoulos2,Contopoulos3}, in the perturbed Schwarzschild spacetime \cite{Bombelli,Bombelli1,Bombelli2,Bombelli3} , in the spacetime of a black hole immersed  an external magnetic field \cite{Karas} and in the accelerating and rotating black holes spacetime \cite{sbch}.
 Moreover,  it is of interest to find that the chaotic behaviors appear in ring string dynamics in the Schwarzschild black hole spacetime \cite{Frolov} and in AdS black hole spacetimes \cite{Zayas,MDZ}.

 The chaotic behavior in van-der Waals fluid \cite{BIAdS5} has been found by applying the Melnikov method \cite{BIAdS6}. The fluid system  is initially supposed to be in the unstable spinodal region, which allows the existence of the homoclinic orbit in phase space. It is shown that for the temporal perturbation the thermal chaos  occurs only if the perturbation amplitude is larger than certain a critical value, but for the spatially perturbation, the chaos always exists irrespective of perturbation amplitude. Recent investigations indicate that thermodynamic behavior of AdS black holes in the extended phase space bears high resemblance to that of van-der Waals fluid system.
 By treating the cosmological constant as a thermodynamic pressure and its conjugate quantity as a thermodynamic volume, Kubiznak et al.\cite{R12} found that in the system of a charged AdS black hole the small-large black hole phase transition possesses the same critical behavior of liquid-gas phase transitions in the van der Waals fluid. The similar $P-V$ critical behaviors are disclosed in the other AdS black hole spacetimes \cite{R120,R121,R122,R123}. These similar critical behaviors mean that there must be some deep connection between AdS black holes and van-der Waals fluid system.
 Thus, it is very natural to probe the thermal chaos in AdS black holes with phase structures similar to that of van-der Waals fluid. For a Reissner-Nordstr\"{o}m-AdS black hole, it is found the similar chaotic behaviors existed in the system of van-der Waals fluid. The critical temporal perturbation amplitude depend on the black hole charge $Q$ and  the presence of  charge makes the onset of chaos easier \cite{BIAdS7}. For a Gauss-Bonnet AdS black hole \cite{BIAdS11}, the presence of charge is necessary for chaos under temporal perturbations. However, under spatial perturbations, chaotic behavior always exist irrespective of whether the black hole carries charge or not. Moreover, the chaos under temporal perturbation is found to occur easier in the case with larger Gauss-Bonnet coupling constant. The same qualitative properties of thermal chaos are also found in the AdS black hole spacetime with Born-Infeld electrodynamics \cite{BIAdS}, which shows that the effect of Born-Infeld parameter on thermal chaos is similar to that of black hole charge.

In this paper, we will study thermal chaos in the extended phase space for a charged dilaton-AdS black hole, which belongs to a family of solutions in  Einstein-Maxwell-dilaton gravity theory \cite{R3ad,R1,R2}. The presence of the dilaton field changes the causal structure of the spacetime and modifies the thermodynamic properties of the black holes.  In an extended phase space, one can find that the thermodynamic quantities depend on the dilaton parameter \cite{R3}.
For example, the thermodynamic pressure $P$ depends on both  the cosmological constant and the coupling strengthen between the dilaton scalar and electromagnetic fields. Correspondingly, the thermodynamic volume $V$ is also a function of the coupling constant, which is different from those in the usual AdS black hole spacetimes in which $P$ is determined entirely by the cosmological constant and $V$ is only a function of the event horizon radius. Although the thermodynamic quantities depend on the dilaton parameter, it is found that the $P-V$ critical behaviors present the similar feature as in usual AdS one and the critical exponents are independent of the details of the dilaton system \cite{R3}. And then, it is very natural to ask whether dilaton parameter affects thermal chaos of AdS black holes in  Einstein-Maxwell-dilaton gravity theory. The main purpose of this paper is to study thermal chaos of a charged dilaton-AdS black hole \cite{R3ad} and to probe the effects of dilaton parameter on thermal chaotic behaviors.

This paper is organized as follows. In Sec.II, we will review briefly thermodynamics of a charged dilaton-AdS black hole \cite{R3ad} in the extended phase space and dependence of the critical temperature $T_c$ on black hole parameters. In Sec. III, we will study the thermal chaos of the charged dilaton-AdS black hole flow under thermal perturbations and probe the effects of dilaton parameter on such kind of thermal chaos. Finally, we present a brief summary.

\section{Thermodynamics of a charged dilaton-AdS black hole in extended phase space}

Let us now review briefly the thermodynamics of a charged dilaton-AdS black hole in extended phase space. In a four-dimensional spacetime, the action of Einstein-Maxwell theory with a dilation field  can be expressed as \cite{R3ad,R1,R2,R3}
\begin{eqnarray}\label{act1}
\mathcal{I}=\frac{1}{16\pi}\int d^{4}x\sqrt{-g}\left[R-2\left(\nabla\varphi\right)^{2}-
2\Lambda e^{2\alpha\varphi}-\frac{2\alpha^{2}}{b^{2}\left(\alpha^{2}-1\right)}
e^{2\varphi/\alpha}-e^{-2\alpha\varphi}F_{\mu\nu}F^{\mu\nu}\right].
\end{eqnarray}
Here the scalar $\varphi$ is dilaton field and the electromagnetic tensor $F_{\mu\nu}=\partial\left[_{\mu}A_{\nu}\right]$ is related to the potential vector $A_{\nu}$.  The quantity $b$ is a positive arbitrary constant and $\alpha$ is the coupling parameter between the dilaton and Maxwell fields.
From the action (\ref{act1}), one can obtain a spherical symmetric black hole solution,  whose metric has the form \cite{R3ad}
\begin{eqnarray}\label{solu1}
ds^{2}&=&-f(r)dt^{2}+\frac{1}{f(r)}dr^{2}+r^{2}R(r)^{2}(d\theta^2+\sin^2\theta d\varphi^2),
\end{eqnarray}
with
\begin{eqnarray}
&&f\left(r\right)=-\frac{\alpha^{2}+1}{\alpha^{2}-1}(\frac{b}{r})^{-2\gamma}
-\frac{m}{r^{1-2\gamma}}-\frac{3\left(\alpha^{2}+1\right)^{2}r^{2}}
{l^{2}\left(\alpha^{2}-3\right)}
\left(\frac{b}{r}\right)^{2\gamma}
+\frac{q^{2}\left(\alpha^{2}+1\right)}{r^{2}}
\left(\frac{b}{r}\right)^{-2\gamma},\\
&& \varphi\left(r\right)=\frac{\alpha}{\alpha^{2}+1}\ln\left(\frac{b}{r}\right),
\quad\quad\quad R(r)=\left(\frac{b}{r}\right)^{\gamma},\quad\quad\quad F_{tr}=\frac{e^{2\alpha\varphi}}{r^2R^2}.
\end{eqnarray}
Here $\gamma$ is related to the coupling parameter $\alpha$ by $\gamma=\frac{\alpha^{2}}{\alpha^{2}+1}$.  The parameter $m$ is associated with the ADM mass $M$ of black hole (\ref{solu1}) by $m= 2(\alpha^{2}+1)b^{-2\gamma}M$ and $q$ is the black hole charge. In the absence of the dilaton field $(\alpha=0)$, the above metric describes the geometry of the usual Reissner-Nordstr\"{om} AdS black hole.

For the black hole (\ref{solu1}), with the event horizon radius $r_+$ (i.e., the largest root of equation $f(r)=0$),  one can obtain Hawking temperature and entropy 
\begin{eqnarray}
&&T=-\frac{\left(\alpha^{2}+1\right)}{4\pi\left(\alpha^{2}-1\right)r_{+}}\left(\frac{b}{r_{+}}\right)^{-2\gamma}-\frac{\varLambda\left(\alpha^{2}+1\right)r_{+}}{4\pi}
\left(\frac{b}{r_{+}}\right)^{2\gamma}-\frac{q^{2}\left(\alpha^{2}+1\right)}{4\pi r_{+}^{3}}\left(\frac{b}{r_{+}}\right)^{-2\gamma},\\
&&S=\pi b^{2\gamma}r_{+}^{2\left(1-\gamma\right)},
\end{eqnarray}
respectively.
Introducing the thermodynamic pressure $P$, its conjugate quantity as volume $V$ and the electric potential $U$
\begin{eqnarray}
P=-\frac{\left(3+\alpha^{2}\right)b^{2\gamma}\Lambda}{8\pi\left(3-\alpha^{2}\right)r_{+}^{2\gamma}},
\quad\quad\quad \quad V=\frac{4\pi\left(1+\alpha^{2}\right)b^{2\gamma}}{3
+\alpha^{2}}r_{+}^{\frac{3+\alpha^{2}}{1+\alpha^{2}}},
\quad\quad\quad \quad U=\frac{q}{r_+},
\end{eqnarray}
one can find that the first law of thermodynamics and the corresponding Smarr formula can be expressed as
\begin{eqnarray}
&&dM=TdS+Udq+VdP,\\
&&M=2(1-\gamma)TS+Uq+\left(4\gamma-2\right)VP,
\end{eqnarray}
respectively.
In the limit $\alpha\rightarrow 0$, the above thermodynamic pressure $P$ and volume $V$ become
\begin{eqnarray}
P=-\frac{\Lambda}{8\pi},\quad\quad\quad\quad  V=\frac{4\pi r_{+}^3}{3},
\end{eqnarray}
which are consistent with that of Reissner-Nordstr\"{om} AdS black hole.
Similarly, defining the specific volume
\begin{eqnarray}
v=\frac{2(1+\alpha^{2})(3-\alpha^{2})}{(3+\alpha^{2})}r_+,
\end{eqnarray}
one can find that thermodynamic pressure $P$ can be rewritten as
\begin{eqnarray}
P(v,T)&=&\frac{T}{v}+\frac{b^{-2\gamma}}{2^{2\gamma}\pi v^{2(1-\gamma)}(\alpha^{2}+1)^{2(\gamma-2)}}\bigg[\frac{\left(3-\alpha^{2}\right)^{1-2\gamma}}{2 (\alpha^{2}-1)(\alpha^{2}+1)^{2}
\left(3+\alpha^{2}\right)^{1-2\gamma}}+\frac{2q^{2}
\left(3+\alpha^{2}\right)^{2\gamma-3}}{ v^{2}\left(3-\alpha^{2}\right)^{2\gamma-3}}\bigg].\label{StateEq}
\end{eqnarray}
\begin{figure}[ht]
\begin{center}
\includegraphics[width=6.2cm]{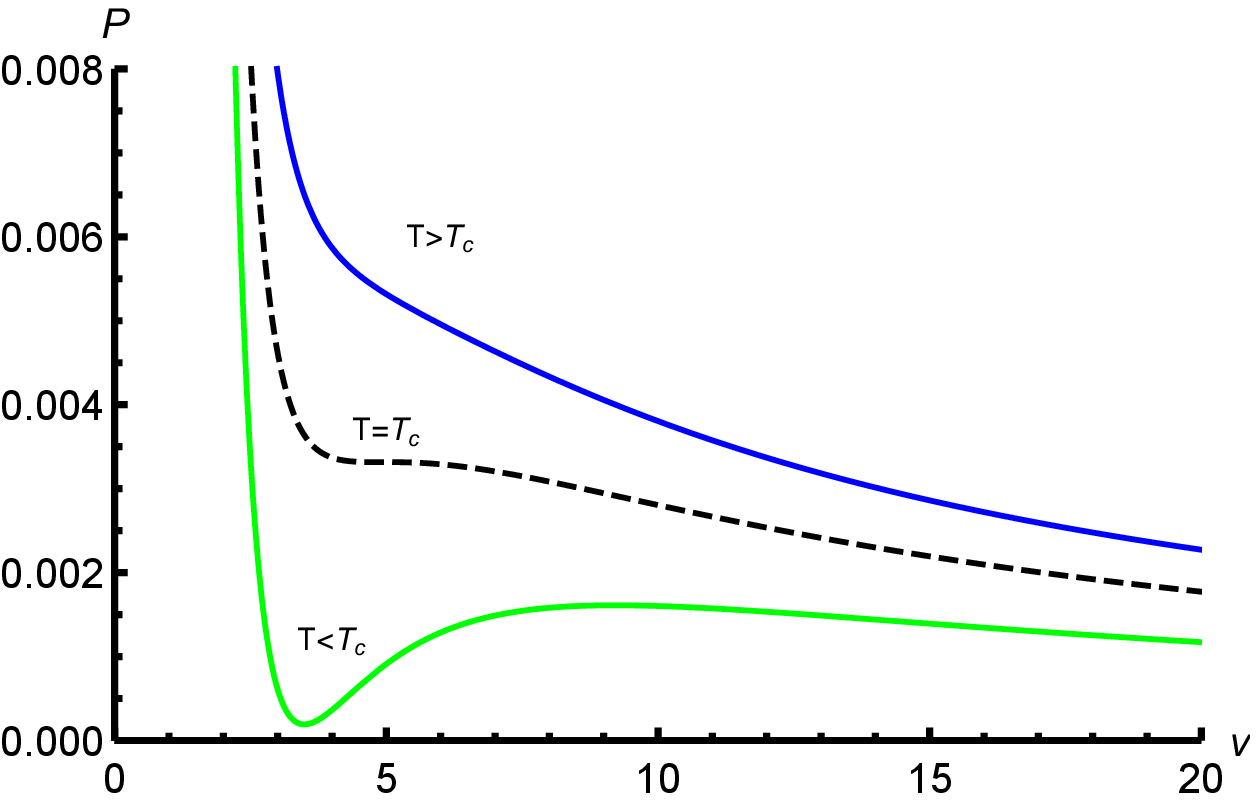}\quad\quad
\includegraphics[width=6.2cm]{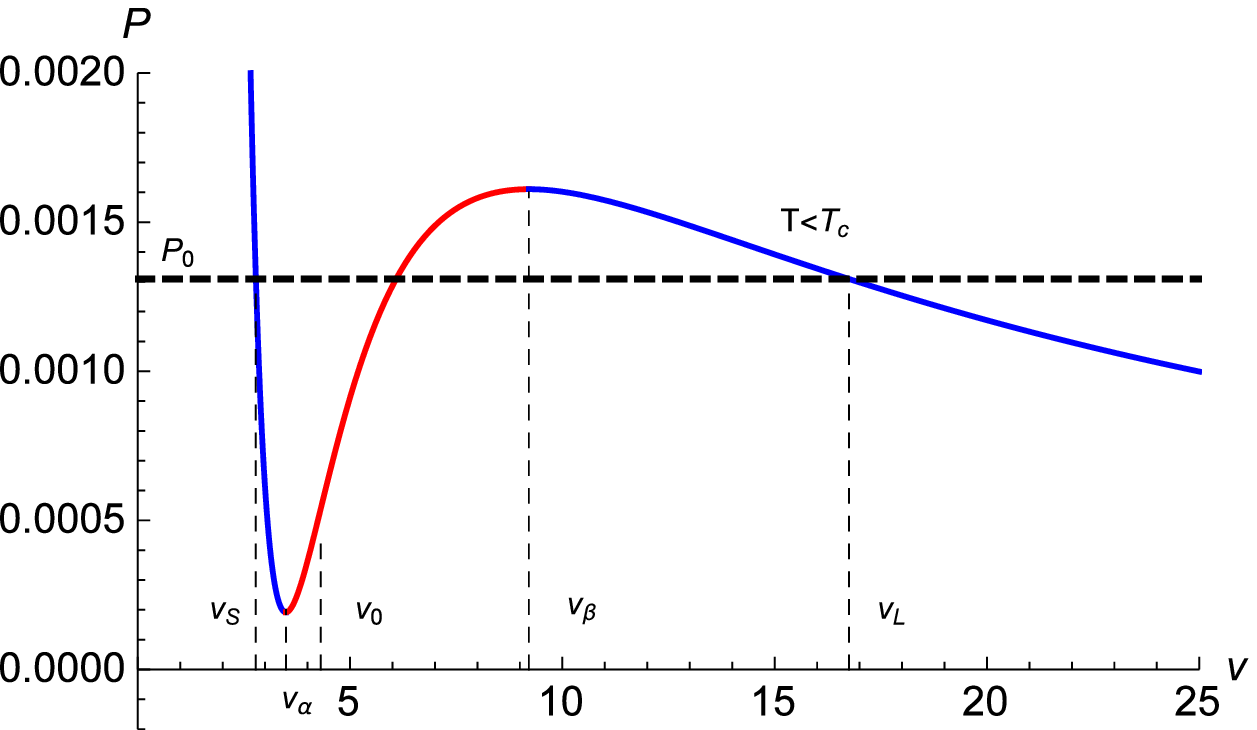}
\caption{ (colour online) The $P-v$ isothermal curves with different temperatures $T$ for the charged dilaton-AdS black holes with the fixed  parameters $b=1$, $q=1$ and $\alpha=0.01$. On the right panel with $T<T_c$, the curve can be divided into two stable regions(blue lines) and a unstable one (red line). The black dashed line is the coexisting line of large black hole (with specific volume $v_L$) and small black hole (with specific volume $v_S$) with phase transition pressure $P_0$. }\label{FIG1}
\end{center}
\end{figure}
The rich phase structures for the black hole (\ref{solu1}) have been studied in the extended phase space \cite{R3}. It is shown in Fig. (\ref{FIG1}) that there exists Large-Small black hole phase transition, which is qualitatively similar to the gas-fluid phase transition in the van-der Waals system. This is a second order phase transition occurred at the critical temperature
 \begin{eqnarray}
T_c=\frac{(\alpha^2+1)(3+\alpha^2)^{\frac{2\gamma-3}{2}}}{\pi b^{2\gamma}q^{1-2\gamma}(1-\alpha^2)(\alpha^2+2)^{\frac{1-2\gamma}{2}}}.\label{tc}
\end{eqnarray}
Obviously, the critical temperature $T_c$ decreases with the parameters $b$ and $q$. The change of $T_c$ with $\alpha$ depends on the value of $q$, which is shown in Fig.(\ref{FIG1s}). The critical temperature $T_c$ for different $b$ increases monotonically with $\alpha$ in the case with the larger $q$, but it first decreases and then increases in the case with the smaller $q$.
The diagrams in Fig. (\ref{FIG1}) indicate that as the temperature of black hole $T$ is less than the critical temperature $T_c$ the $P-v$ curve can be divided into two stable regions and a unstable one. Two stable regions $v \in [0,v_\alpha]$ and $v \in [v_\beta, \infty]$ correspond to the small black hole region and the large black hole one, respectively. The two points $v_{\alpha}$ and $v_{\beta}$ are determined by $\frac{\partial P(v,T_0)}{\partial v}|_{v_{\alpha}}=\frac{\partial P(v,T_0)}{\partial v}|_{v_{\beta}}=0$. The unstable region $v \in [v_\alpha, v_\beta]$ is the so-called spinodal region in which the small and large
black hole phase coexist. Obviously, one can find that $\frac{\partial P(v,T_0)}{\partial v}<0$ in the stable regions, but $\frac{\partial P(v,T_0)}{\partial v}>0$ in the unstable one. Here, $v_\alpha$ and $v_\beta$ are two extreme points, which satisfy the equation  $\frac{\partial P(v,T_0)}{\partial v}=0$. The inflection point at $v=v_0$ is determined by $\frac{\partial^2 P}{\partial^2 v}=0$.
\begin{figure}[ht]
\begin{center}
\includegraphics[width=5.2cm]{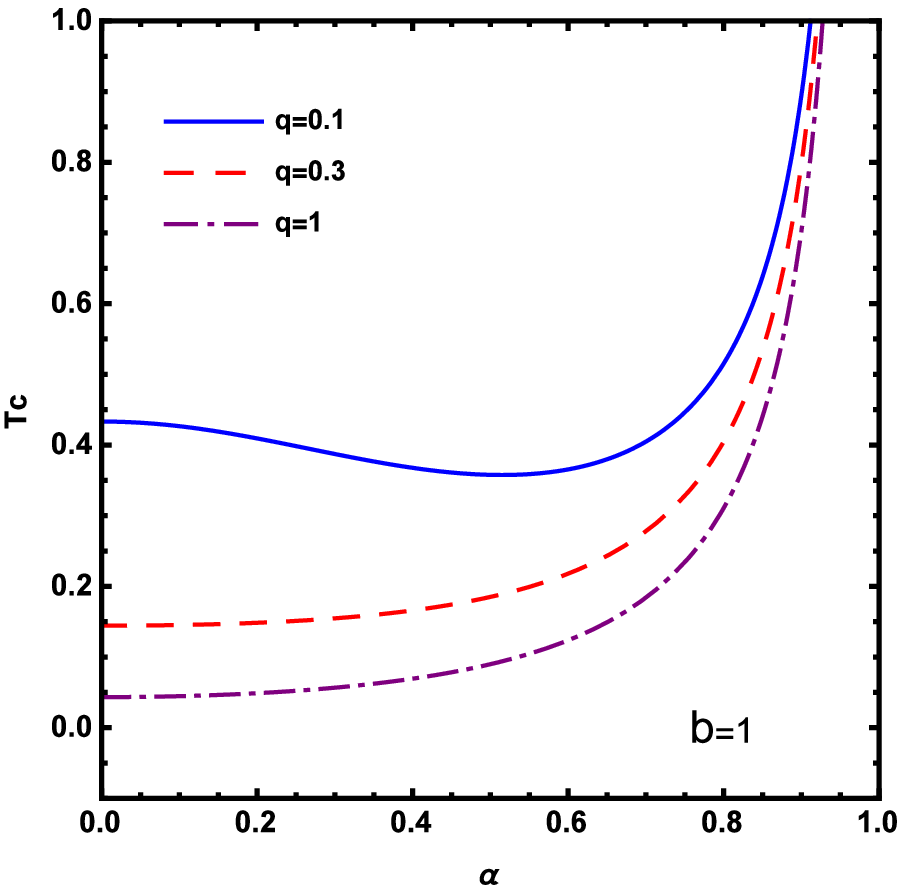}\quad\includegraphics[width=5.2cm]{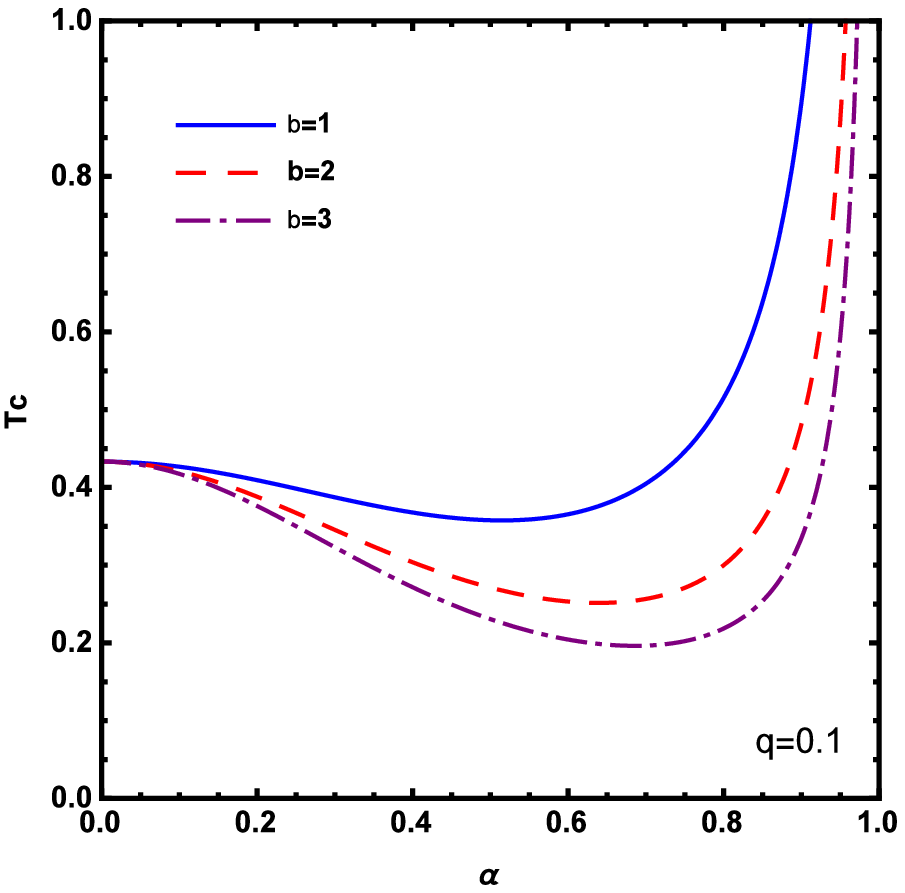}
\quad\includegraphics[width=5.2cm]{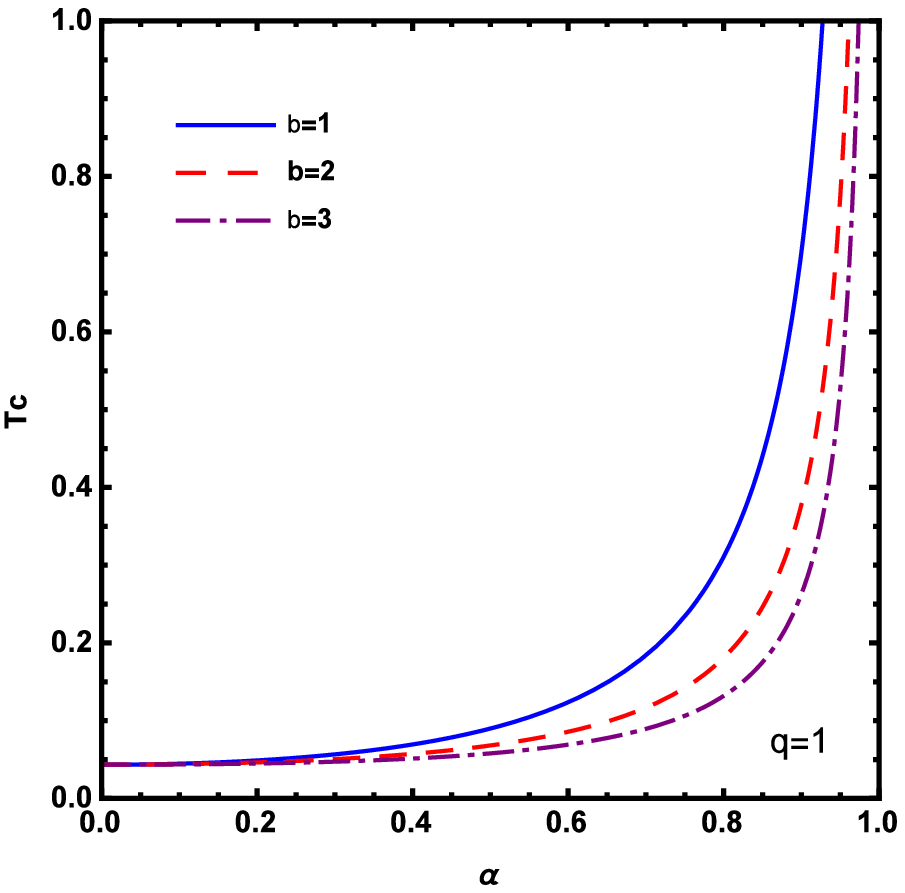}
\caption{ (colour online) Dependence of the critical temperature $T_c$ on parameter $\alpha$ for the charged dilaton-AdS black hole with different $b$ and $q$.} \label{FIG1s}
\end{center}
\end{figure}
As in refs.\cite{BIAdS5,BIAdS7,BIAdS,BIAdS11}, we will study the chaotic behaviour of the charged dilaton-AdS black hole under periodic thermal perturbations through the Melnikov method.

\section{Chaos in the charged dilaton-AdS black hole flow under thermal perturbations}

As in refs.\cite{BIAdS5,BIAdS7,BIAdS,BIAdS11}, for a sake of simplicity, the charged dilaton-AdS black hole flow is assume to move along  $x$-axis in a tube of unit cross section of fixed volume. And then, the position of a fluid particle can be described  by the Eulerian coordinate $x$. The mass $M$ of a column of fluid of unit cross section between the reference fluid particle with Eulerian coordinate  $x_\ast$ and a fluid particle with the coordinate $x$ can be expressed as
\begin{eqnarray}\label{mx}
M=\int^{x}_{x_\ast}\rho(\xi,t)d\xi,
\end{eqnarray}
where $\rho(x,t)$ is the fluid density at spatial position $x$ and time $t$. The relation (\ref{mx}) means that the position $x$ of any particle can be expressed as a function of the mass $M$ and the time $t$, i.e., $x=x(M,t)$.
From Eq.(\ref{mx}), it is easy to get $x_M(M,t) \equiv \frac{\partial x(M,t)}{\partial M}=\rho[x(M,t),t]^{-1}$, which is defined as the specific volume $v(M,t)$. Similarly, one can also define the velocity as $u(M,t) \equiv x_t (M,t)$. With these quantities, one can find that the charged dilaton-AdS black hole flow can be described by a dynamical system with the balance equation of mass and momentum \cite{BIAdS5,BIAdS7,BIAdS,BIAdS11}
\begin{eqnarray}
 \frac{\partial v}{\partial t} &=& \frac{\partial u}{\partial M}, \\ \frac{\partial u}{\partial t} &=& \frac{\partial \tau}{\partial M},\label{MomentumBalance}
\end{eqnarray}
 where $\tau$ is the Piola stress. As in refs.\cite{BIAdS5,BIAdS7,BIAdS,BIAdS11}, we assume that the charged dilaton-AdS black hole fluid is thermoelastic, isotropic and slightly viscous, which means that the Piola stress can be expressed as \cite{BIAdS10}
 \begin{eqnarray}\label{tao1}
 \tau = -P(v,T)+\mu u_M-A v_{MM}.
\end{eqnarray}
Here $A$ is a positive constant, and $\mu$ is a small positive constant viscosity. With the above form of Piola stress $\tau$, the balance equation (\ref{MomentumBalance}) can be rewritten as
 \begin{eqnarray}\label{xtt0}
 x_{tt}=-P(v,T)_M+\mu u_{MM}-Av_{MMM}.
\end{eqnarray}
One can suppose that a finite black hole fluid tube of unit cross section contains the total mass $2\pi/s$ in a volume $2\pi v_0/s$, where $s$ is a positive parameter. Introducing the change of variables $\tilde {M}=s M, \tilde{t}=s t, \tilde{x}=s x$ and $\mu=\epsilon \mu_0$, one can find that the range of the mass $\tilde {M}$  becomes $\tilde{M} \in [0, 2\pi]$,  and then the equation (\ref{xtt0}) can be written as
 \begin{eqnarray}
 x_{tt}=-P(v,T)_M+\epsilon \mu_0 s u_{MM}-A s^2v_{MMM},\label{MainEq}
\end{eqnarray}
where $\epsilon $ is a perturbation parameter. Here the overbars are omitted for later convenience.

\begin{figure}[ht]
\begin{center}
\includegraphics[width=6cm]{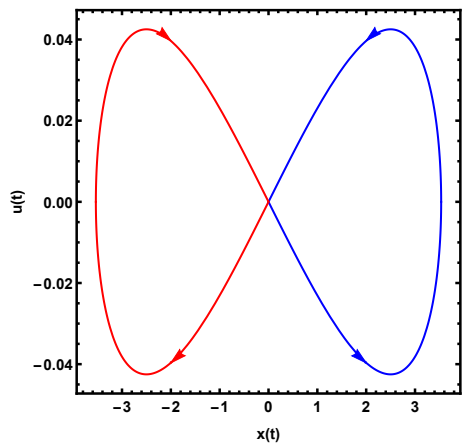}
\caption{ (colour online) Homoclinic orbit of the unperturbed equations for the charged dilaton-AdS black hole with $T=0.0315<T_c$.  The arrows denote the time flow direction. Here, we set $ q=1$ , $ b=1$, $ A=0.2$, $s=0.001$, and $\alpha=0.01$.}\label{FIG2}
\end{center}
\end{figure}

\begin{figure}[ht]
\begin{center}
\subfigure[$\delta=0.000015<\delta_c$]{\includegraphics[width=6cm]{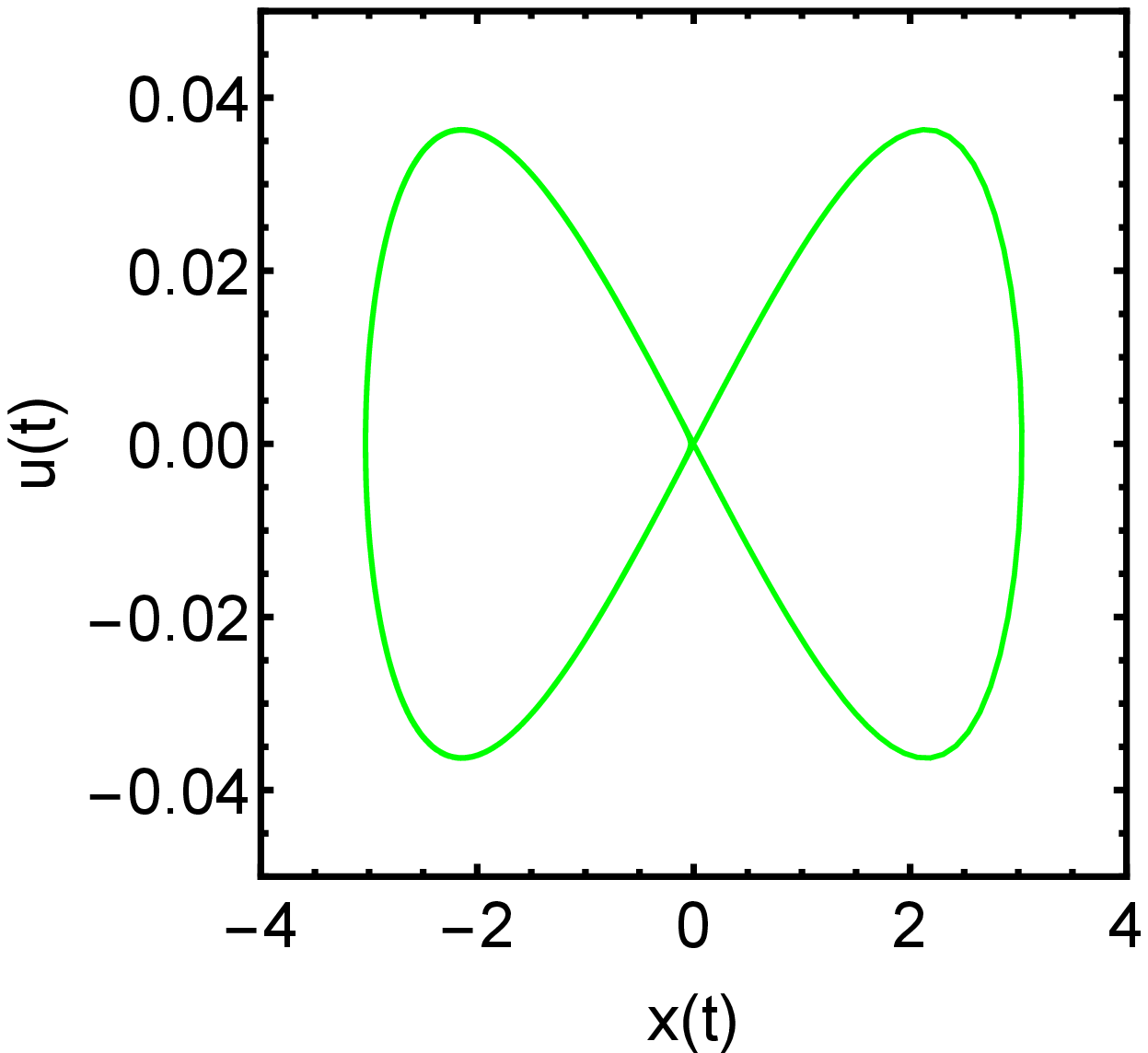}}
\subfigure[$\delta=5>\delta_c$]{\includegraphics[width=6cm]{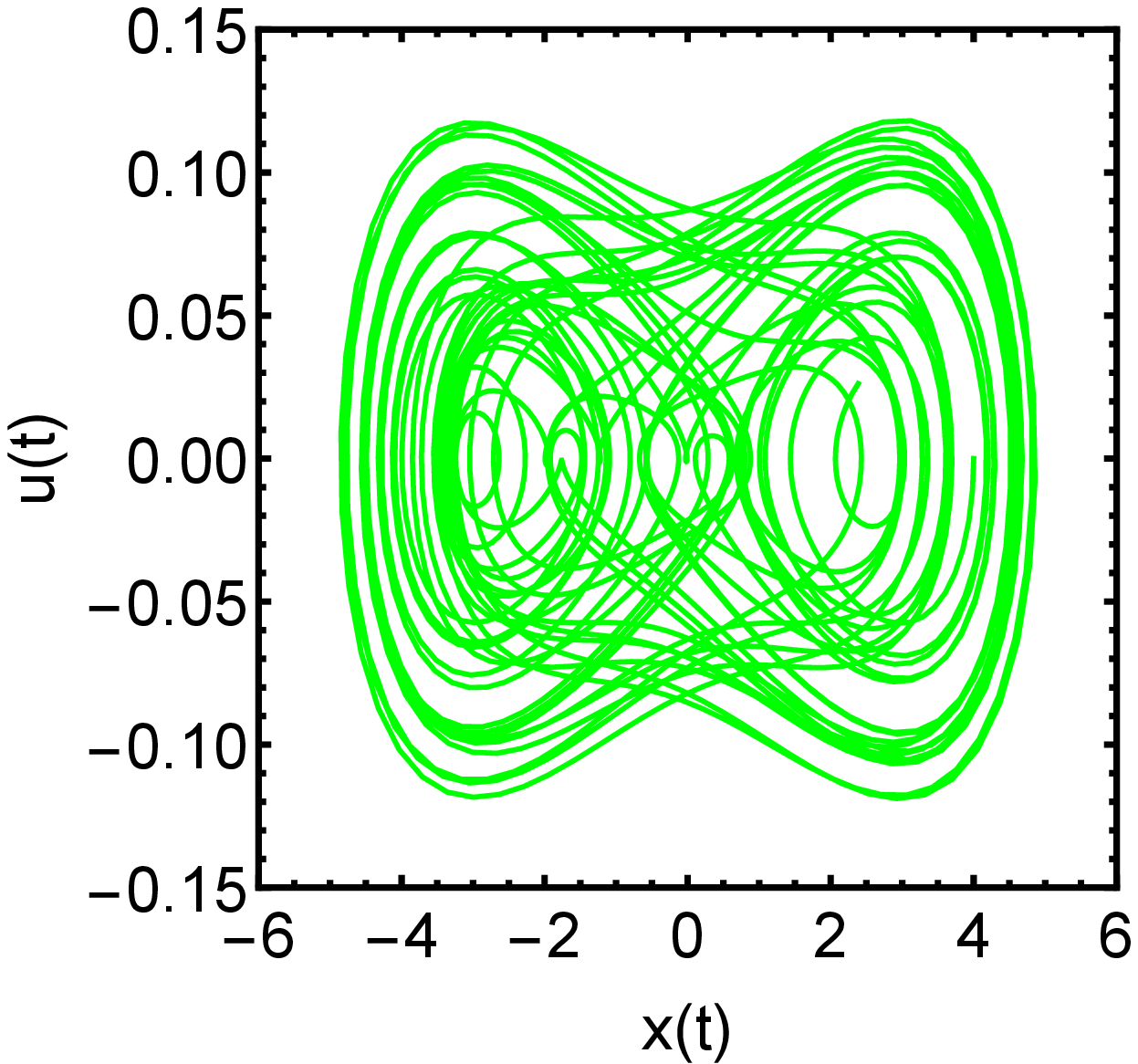}}
\caption{ (colour online) Evolution of the perturbed equations with time $t$ in $x-u$ plane for the fixed system temperature $T=0.0313<T_c$. Parameters are fixed as $\omega=0.01, \epsilon=0.001, \mu_0 = 0.1 $ with other parameters set as in Fig.\ref{FIG2}. The critical value $\delta_c \approx 0.0000208$. }\label{FIG3}
\end{center}
\end{figure}

Let us now to study the temporal chaos in the spinodal region in which the small and large
black hole phase coexist. The system is assumed to lie initially in the unstable equilibrium state associated with the specific volume $v=v_0$ (the inflection point) and the temperature $T=T_0<T_c$. The small time-periodic fluctuation of
the absolute temperature about $T_0$ has a form \cite{BIAdS5,BIAdS6,BIAdS7,BIAdS11,BIAdS}
\begin{eqnarray}\label{Tper}
T=T_0 +\epsilon \delta \cos{\omega t}\cos M,
\end{eqnarray}
where $\delta$ is amplitude of the perturbation relative to the small viscosity. Expanding $P(v,T)$ around the equilibrium point $(v_0,T_0)$ as in refs.\cite{BIAdS5,BIAdS6,BIAdS7,BIAdS11,BIAdS}, we have
\begin{eqnarray}
P(v,T)&=&P(v_0,T_0)+P_v(v_0,T_0)(v-v_0)+P_T(v_0,T_0)(T-T_0)+P_{vT}(v_0,T_0)(v-v_0)(T-T_0) \nonumber \\&& +\frac{P_{vvv}(v_0,T_0)}{3!}(v-v_0)^3+\frac{P_{vvT}(v_0,T_0)}{2}(v-v_0)^2(T-T_0) +\cdots,\label{PExpansion}
\end{eqnarray}
with
\begin{eqnarray}
   P_{T}(v_0,T_0)&=&\frac{1}{v_0},\quad\quad\quad\quad
   P_{vT}(v_0,T_0)=-\frac{1}{v^2_0},\quad\quad\quad\quad
   P_{vvT}(v_0,T_0)=\frac{2}{v^3_0},\nonumber\\
P_v(v_0,T_0)&=&-\frac{T_0}{v^2_0}+\frac{\left[(3-\alpha^2)(\alpha
   ^2+1)\right]^{\frac{1-\alpha ^2}{\alpha
   ^2+1}}\left[4(\alpha^4+\alpha^2-2)(\alpha^4-2\alpha^2-3)^2 q^2+\left(\alpha ^2+3\right)^2
   v^2_0\right]}{\pi (1-\alpha ^2)4^{\frac{\alpha ^2}{\alpha ^2+1}}b^{\frac{2 \alpha ^2}{\alpha
   ^2+1}}
   \left(\alpha ^2+3\right)^{\frac{\alpha
   ^2+3}{\alpha ^2+1}} v_0^{\frac{3 \alpha ^2+5}{\alpha ^2+1}}},\nonumber
\end{eqnarray}
\begin{eqnarray}
P_{vvv}(v_0,T_0)&=&-\frac{6T_0}{v^4_0}+\frac{2\left(\alpha^2-3\right)^{\frac{1-\alpha
   ^2}{1+\alpha ^2}}(\alpha^2+2)}{\pi
    \left(1-\alpha ^2\right) 4^{\frac{\alpha
   ^2}{1+\alpha ^2}} b^{\frac{2 \alpha
   ^2}{1+\alpha ^2}}\left(1+\alpha
   ^2\right)^{\frac{3\alpha^2+1}{1+\alpha ^2}} \left(\alpha
   ^2+3\right)^{\frac{\alpha ^2+3}{1+\alpha ^2}}
   v_0^{\frac{5\alpha^2+7}{\alpha^2+1}}}\bigg[(\alpha^2+3)^3v_0^2,\nonumber\\
   &+&2(\alpha^2-1)(3\alpha^2+5)(5\alpha^2+7)
(\alpha^4-2\alpha^2-3)^2q^2\bigg].
\end{eqnarray}
The coefficients $P_{TT}(v_0,T_0)$, $P_{vTT}(v_0,T_0)$ and $P_{TTT}(v_0,T_0)$ vanish because the pressure (\ref{StateEq}) is a linear function of the temperature $T$. The absence of $P_{vv}(v_0,T_0)$ is attributed to a fact that the thermodynamics system (\ref{StateEq}) satisfies $\frac{\partial^2 P(v,T_0)}{\partial v^2}=0$ at the inflection point $v=v_0$.
Moreover, near the inflection point,  the functions $v(M,t)$ and $u(M,t)$ can be expanded in Fourier sine and cosine series on $M\in [0,2\pi]$ respectively, i.e.,\cite{BIAdS5,BIAdS6,BIAdS7,BIAdS11,BIAdS}
\begin{eqnarray}
v(M,t)&=&x_M(M,t)=v_0+x_1 (t)\cos M + x_2 (t) \cos 2M + x_3 (t) \cos 3M + \cdots,\nonumber \\
u(x,t)&=&x_t(M,t)=u_1(t)\sin M + u_2 (t) \sin 2M + u_3 (t) \sin 3M + \cdots.\label{ModesExpansion}
\end{eqnarray}
Here $x_i (t)$ and $u_i(t)$ can be regarded as hydrodynamical modes which describe the deviation from the initial equilibrium state with $v=v_0$. Although a full analysis should consider the effects from the higher modes, it is very difficult to find  the homoclinic orbit in the case containing the higher modes, even in the three-mode one \cite{BIAdS5}. It is argued that the existence of homoclinic orbits for higher mode approximations to the original partial differential equation or for the
infinite-dimensional problem itself remains an open problem \cite{BIAdS5}. The two-mode case is investigated for the van der Waals fluid \cite{BIAdS5} and for charged AdS black hole \cite{BIAdS7} and Gauss-Bonnet one \cite{BIAdS11}.
The recent investigations \cite{BIAdS} show that the effects of the second mode has no basic difference from that of the first mode. Therefore, we here consider only the first mode ($x_1(t)$, $u_1(t)$ ) as in ref.\cite{BIAdS} and omit the subscript $1$ in the later formulae. In doing so, the dynamical equation (\ref{MainEq}) can be simplified further as
\begin{eqnarray}\label{xut}
\dot{x}&=&u, \nonumber \\
\dot{u}&=&(P_v-A s^2)x+\epsilon \left(P_T+\frac{3P_{vvT}}{8}x^2 \right)\delta \cos{\omega t}+\frac{P_{vvv}}{8}x^3-\epsilon \mu_0 s u.
\end{eqnarray}
With the denotation $z \equiv [x, u]^T$, the above equations can be rewritten as
\begin{eqnarray}
 \dot{z}=f(z)+\epsilon g(z,t), \label{GeneralEq}
\end{eqnarray}
with
\begin{eqnarray}
	f(z)=
	\left [
	\begin{array}{c}
		u \\\\
		a^2 x +\frac{P_{vvv}}{8} x^3
	\end{array}
	\right ], \label{fFunction}
\end{eqnarray}
and
\begin{eqnarray}
	g(z)=
	\left [
	\begin{array}{c}
		0 \\\\
		\left( P_T+\frac{3 P_{vvT}}{8} x^2 \right)\delta \cos {\omega t}-\mu_0 s u
	\end{array}
	\right ],\label{gFunction}
\end{eqnarray}
where $a^2 \equiv (P_v-A s^2)$. In the case without thermal perturbation (i.e., $\epsilon=0$), one can find an analytical solution for the equation (\ref{GeneralEq}) \cite{BIAdS,BIAdS12}
\begin{eqnarray}
	z_0(t)=
	\left [
	\begin{array}{c}
		x_0(t) \\\\
		u_0(t)
	\end{array}
	\right ]=
	\left [
	\begin{array}{c}
		\frac{ \pm 4a}{(-P_{vvv})^{1/2}} \text{sech} (at) \\\\
		\frac{\mp 4a^2}{(-P_{vvv})^{1/2}} \text{sech} (at) \tanh (at)
	\end{array}
	\right ]. \label{HomoclinicOrbit}
\end{eqnarray}
It is a homoclinic orbit which joins a saddle equilibrium point to itself. The solution (\ref{HomoclinicOrbit}) owns the two branches, which correspond to two wings of the butterfly-like orbit shown in Fig.(\ref{FIG2}), respectively.
Under the time-periodic thermal perturbation (\ref{Tper}) (i.e.,$\epsilon\neq0$), the above homoclinic orbit may break so that the possible chaos could appear in this system. The existence of chaos is determined by the Melnikov function, which has a form \cite{BIAdS8,BIAdS9}
\begin{eqnarray}
M\left(t_{0}\right)=\int_{-\infty}^{+\infty}f^{\mathrm{T}}\left[z_{0}
\left(t-t_{0}\right)\right]\mathbf{J}g\left[z_{0}\left(t-t_{0}\right)\right]dt,\label{MelnikovFunc}
\end{eqnarray}
with
\begin{eqnarray}
\mathbf{J}=\left[\begin{array}{cc}
0 & 1\\
-1 & 0
\end{array}\right].
\end{eqnarray}
Combining Eq. (\ref{MelnikovFunc}) with Eqs.(\ref{fFunction}) and (\ref{gFunction}), one can obtain the Melnikov function for
the charged dilaton-AdS black hole flow
\begin{eqnarray}
M(t_0)&=& \int_{-\infty}^\infty dt \frac{4a^2}{(-P_{vvv})^{1/2}}\frac{\delta \cos \omega t \sinh a(t-t_0)}{ \cosh^2a(t-t_0)} \left[\frac{6a^2P_{vvT}}{P_{vvv}} \text{sech}^2 a(t-t_0)-P_T \right]\nonumber \\&&+\int_{-\infty}^\infty dt \frac{16\mu_0 s a^4}{P_{vvv}} \text{sech}^2 a(t-t_0) \tanh^2 a(t-t_0).
\end{eqnarray}
With the residue theorem, the Melnikov function can be further expressed as
\begin{eqnarray}
M(t_0)=\delta \omega K \sin \omega t_0+\mu_0 s L,
\end{eqnarray}
with
\begin{eqnarray}
K=\frac{8 \pi}{(-P_{vvv})^{1/2}} \left[P_T-\frac{P_{vvT}}{P_{vvv}}(\omega^2+a^2) \right]\frac{\exp(\frac{\pi \omega}{2a})}{1+\exp(\frac{\pi \omega}{a})},\qquad
L=\frac{32a^3}{3P_{vvv}}.
\end{eqnarray}
The coefficients $K$ and $L$ are similar to those obtained in ref.\cite{BIAdS}, but they depend on the dilaton parameter $\alpha$ in this case. After a simple analysis, one can find that $M(t_0)$ has a simple zero  if the condition
\begin{eqnarray}
\left|\frac{s\mu_0L}{\delta \omega K}\right|\leq1,
\end{eqnarray}
is satisfied \cite{homo chaos}. This means that the chaos appears if the amplitude of the perturbation $\delta$ is larger than $\delta_c=\left|\frac{s\mu_0L}{\omega K}\right|$. In Fig.(\ref{FIG3}), we present numerically the time evolution of equations of motion (\ref{xut}) in the phase plane $x(t)-u(t)$ for the charged dilaton-AdS black hole. It is shown that as $\delta<\delta_c$, the perturbation decays with time and the system will finally approach to its original unstable equilibrium state with $v=v_0$. As $\delta>\delta_c$, one can find that the trajectories in the phase plane become irregular and complex, and  the dynamical evolution of the system exhibits chaotic feature.
\begin{figure}[ht]
\begin{center}
\includegraphics[width=5.3cm]{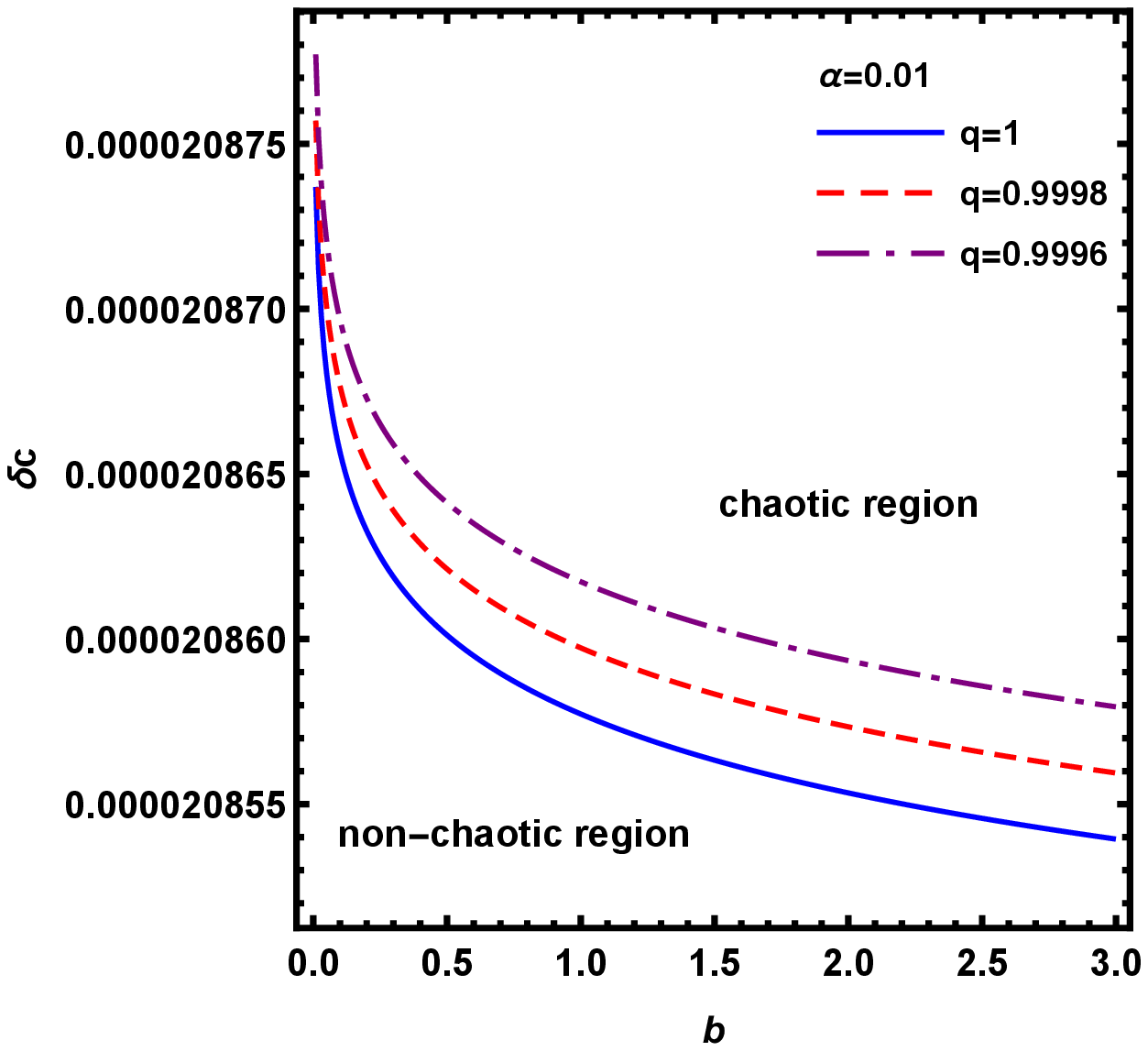}\includegraphics[width=5.3cm]{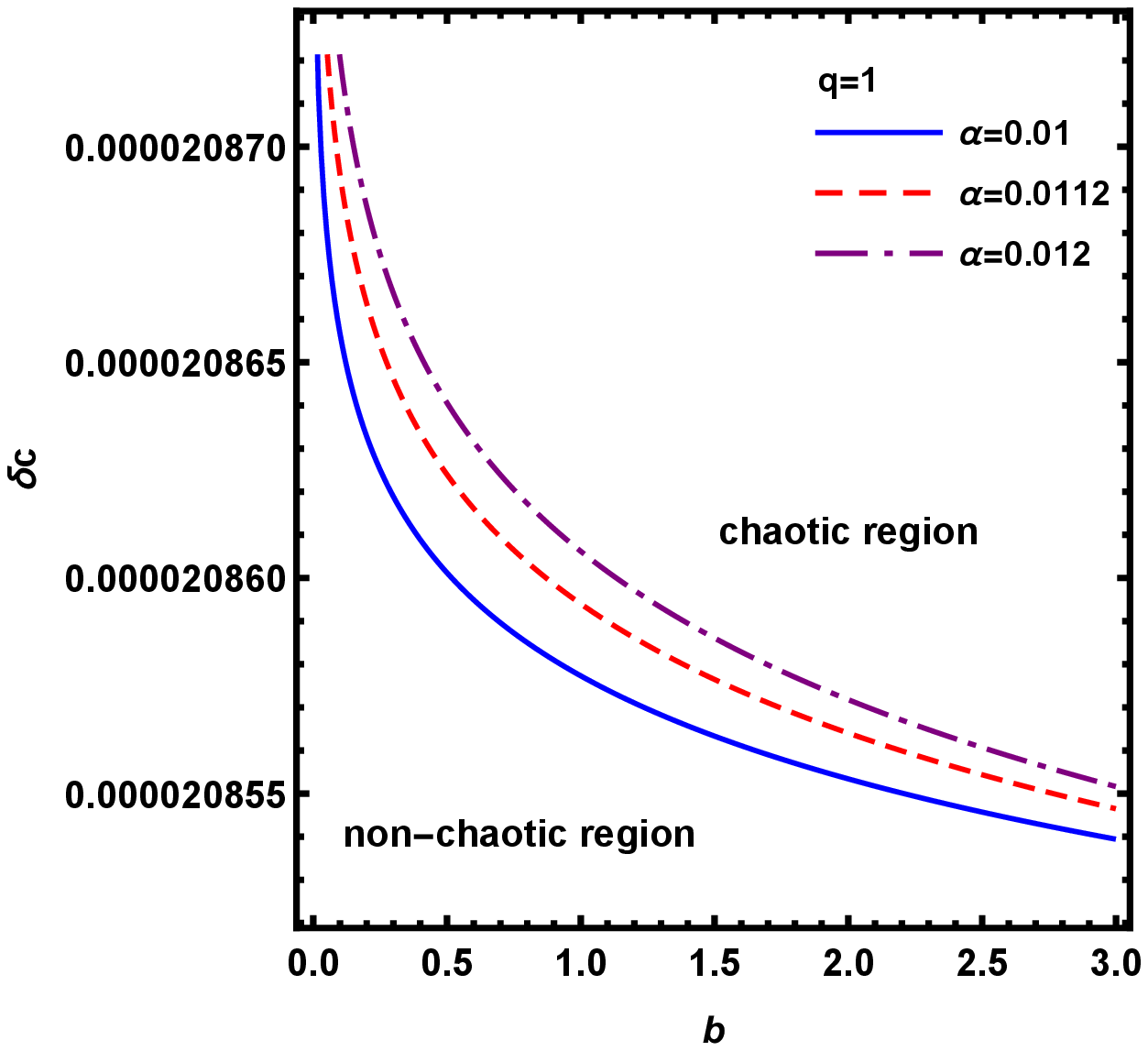}
\includegraphics[width=5cm]{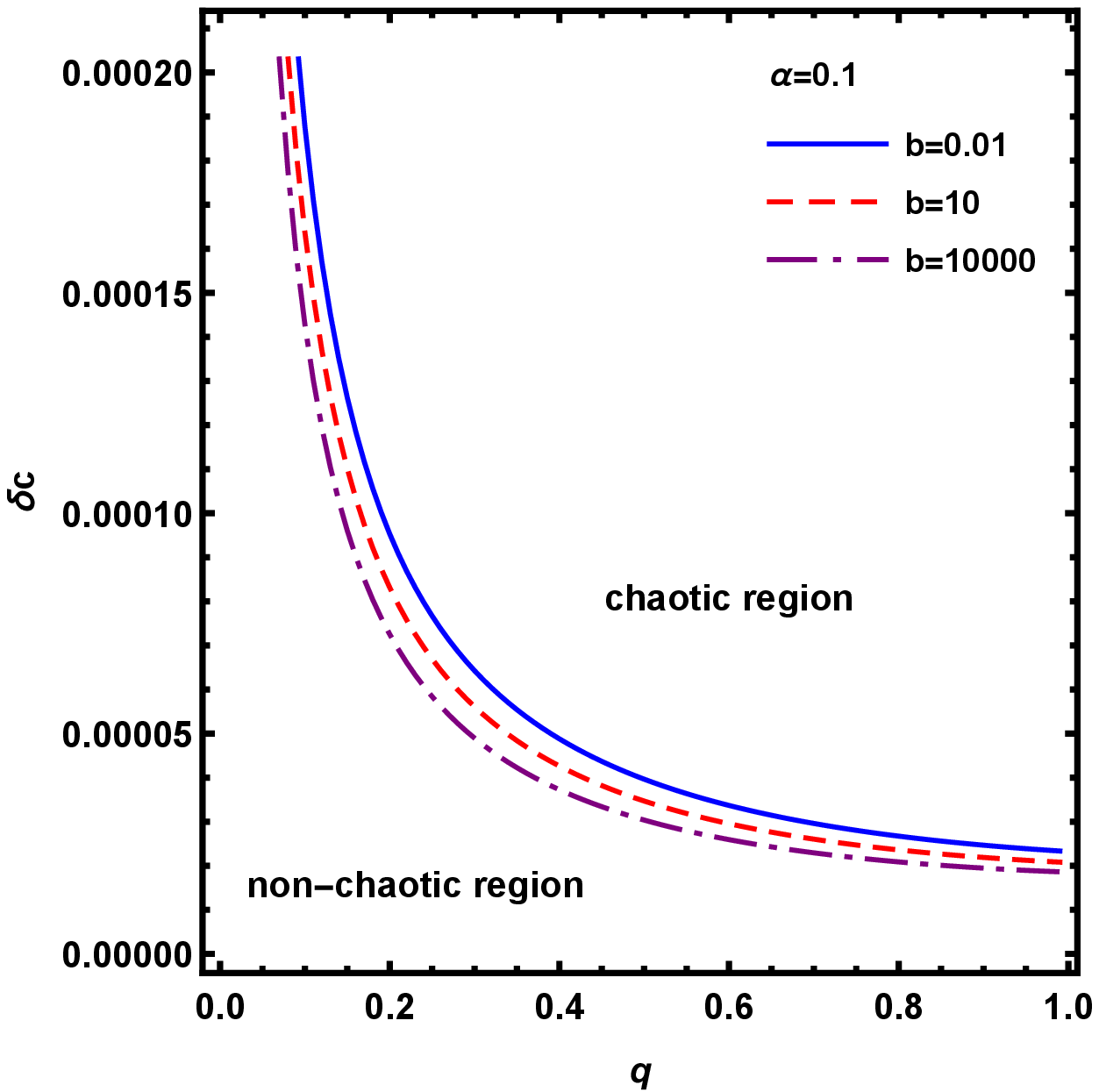}\\\includegraphics[width=4.8cm]{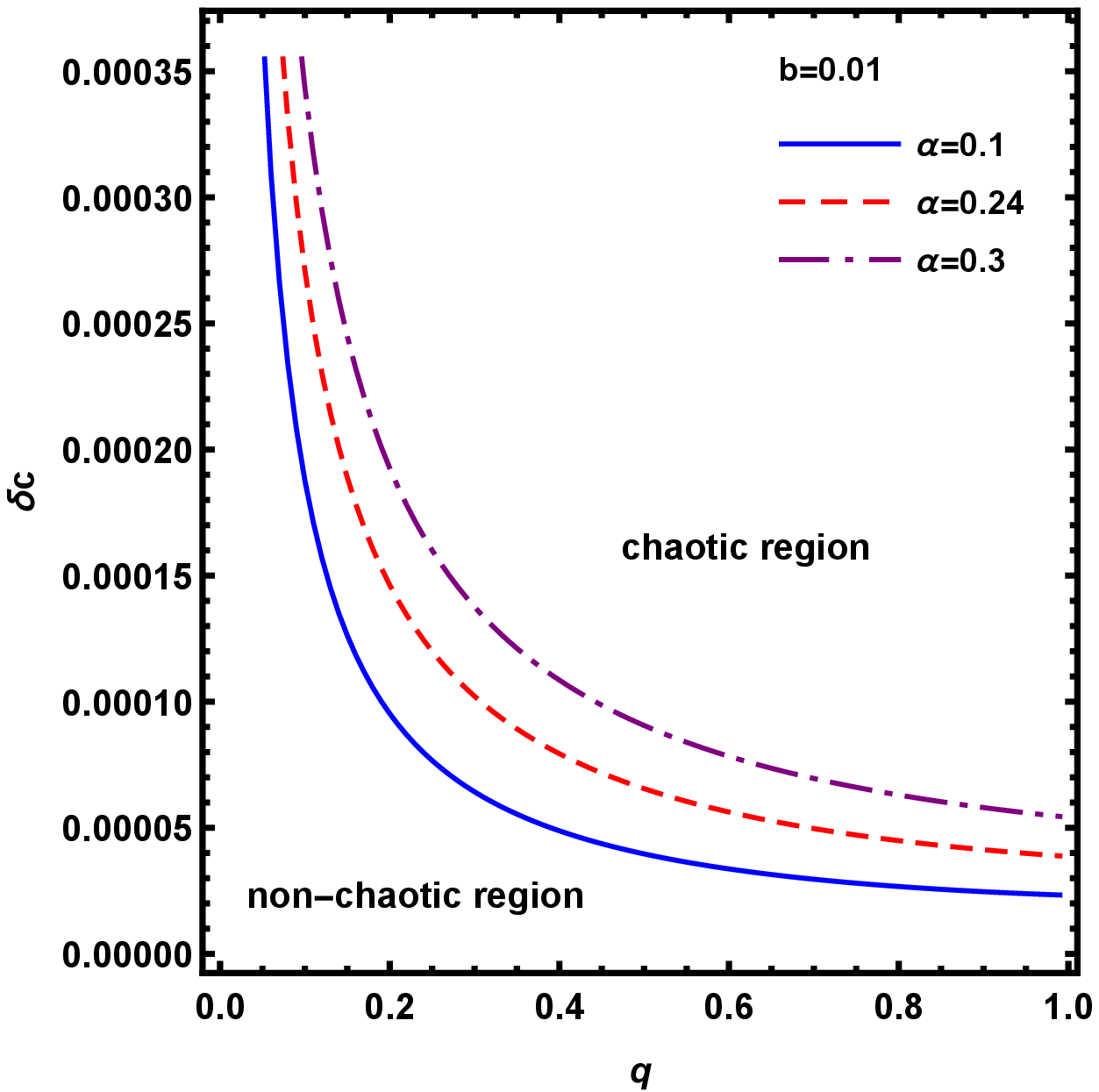}
\includegraphics[width=5cm]{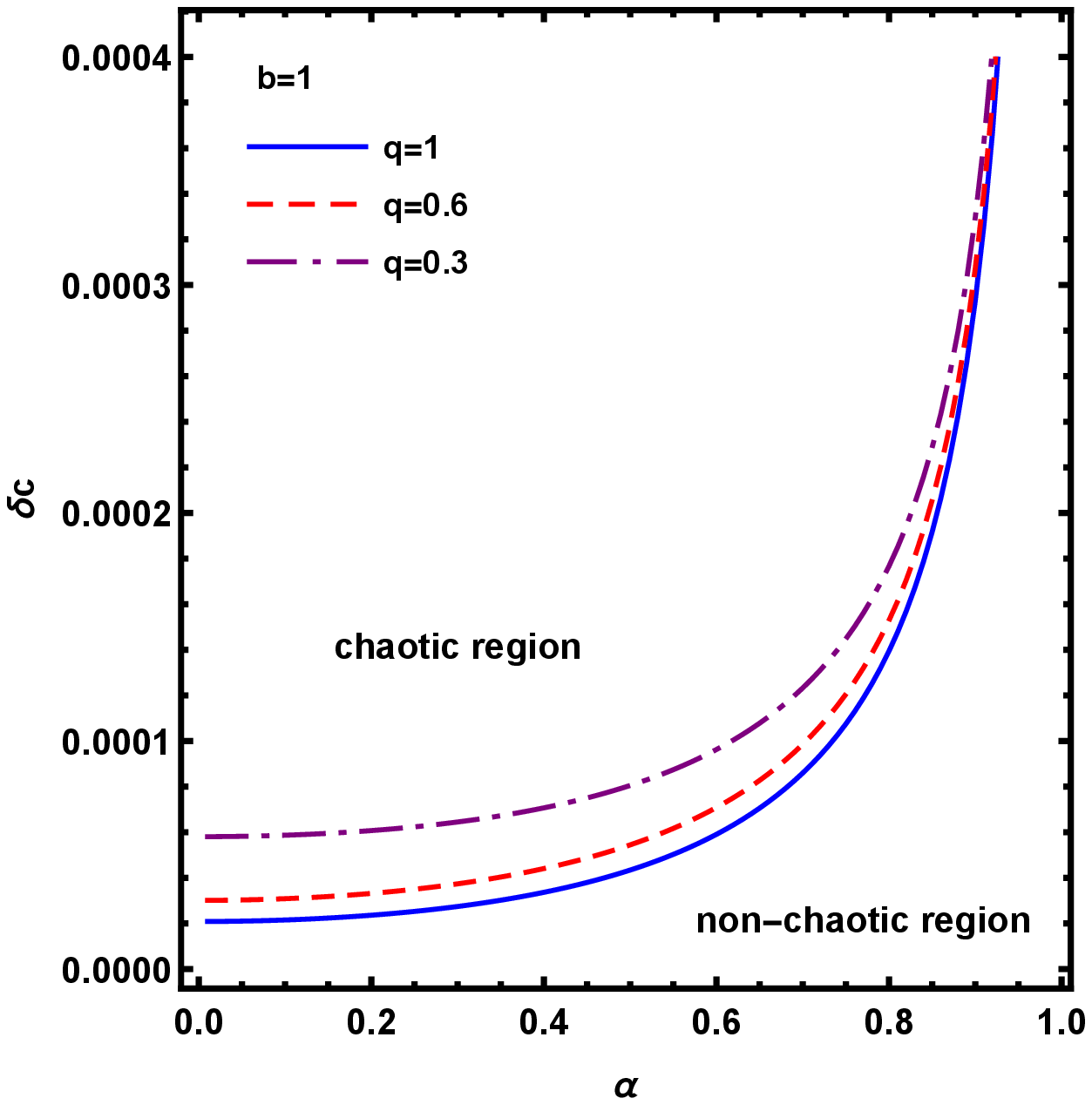}\includegraphics[width=5cm]{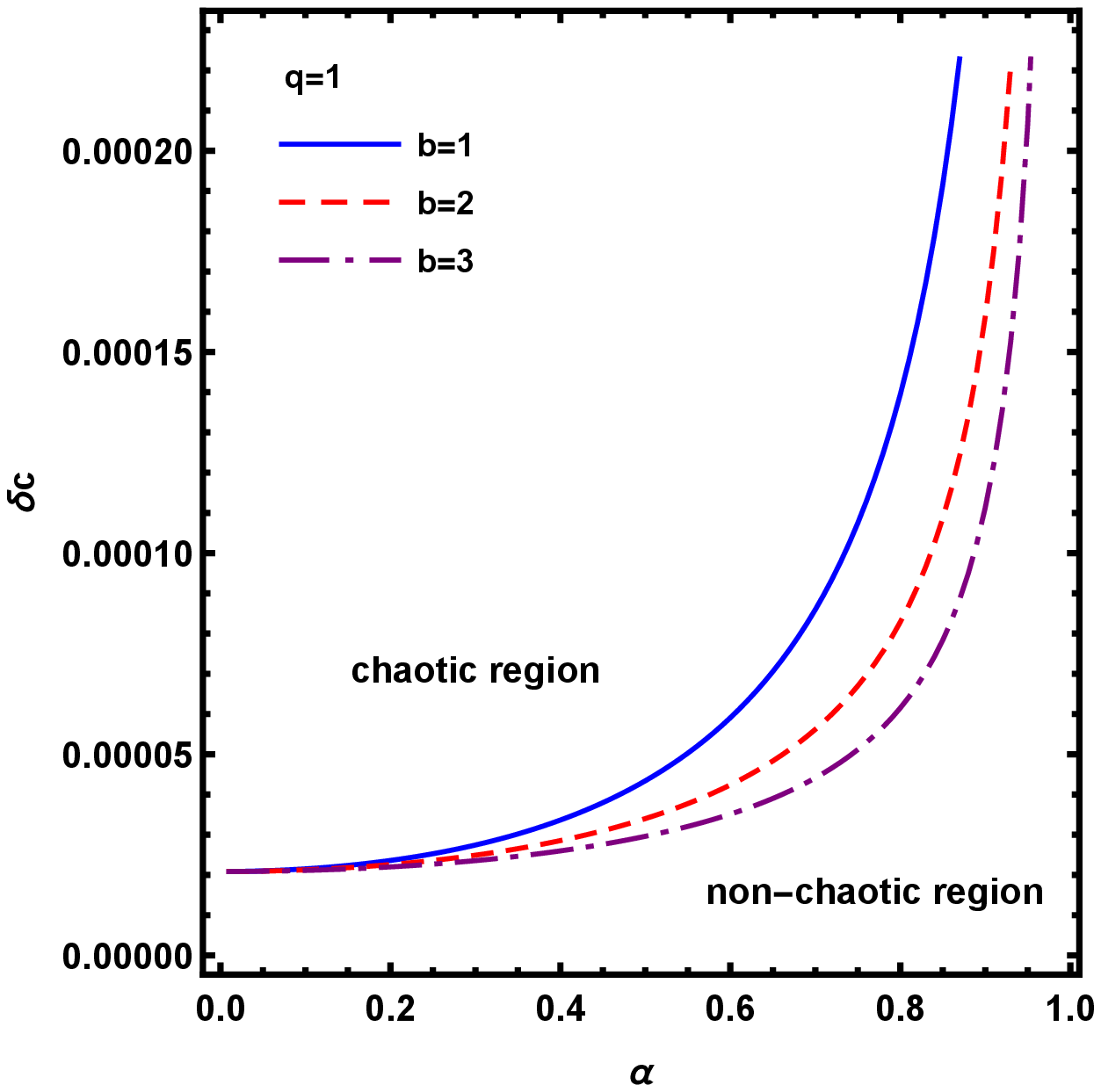}
\caption{ (colour online) Dependence of the critical value $\delta_c$ on the parameters $q$, $b$ and $\alpha$ for the charged dilaton-AdS black hole. Here, we set $T_0/T_c=0.722<1$ to ensure that the system lies in the spinodal region. Other parameters are set as in Fig.\ref{FIG2}.}\label{FIG14}
\end{center}
\end{figure}
The critical value $\delta_c$ depends on various the parameters $b$, $q$ and $\alpha$. From Fig.(\ref{FIG14}), we find that $\delta_c$ decreases with the charge $q$ and the parameter $b$, but increases with the dilaton parameter $\alpha$. This means that the larger $b$ or $q$ makes the onset of chaos easier, but the larger $\alpha$ makes chaos more difficult under the temporal perturbation.

We are now in the position to study the thermal chaos a charged dilaton-AdS black hole due to a small spatially periodic perturbation. Firstly, we assume that the black hole is in the equilibrium state with a sub-critical temperature $T_0$. From van-der Waals-Korteweg theory, the stress tensor without
flow (\ref{tao1}) becomes
\begin{eqnarray}
\tau =-P(v,T_0)-A v^{''},\label{Stress}
\end{eqnarray}
where the notation $'$ denotes the derivative with respect to $x$. $P$ is thermodynamic pressure (\ref{StateEq}) of the charged dilaton-AdS black hole and $A$ is a positive constant. For a static equilibrium state with no body forces, the balance of linear momentum is $\tau'=0$, which means that $\tau =B=constant$. This constant quantity $B$ also represents the ambient pressure at the end of the tube. Thus,  the equation (\ref{Stress}) for a static equilibrium state can be expressed further as
\begin{eqnarray}
A v''+P(v,T_0)=B, \quad\quad \quad  -\infty<x<\infty. \label{MainEq2}
\end{eqnarray}
\begin{figure}[ht]
\begin{center}
\includegraphics[width=5cm]{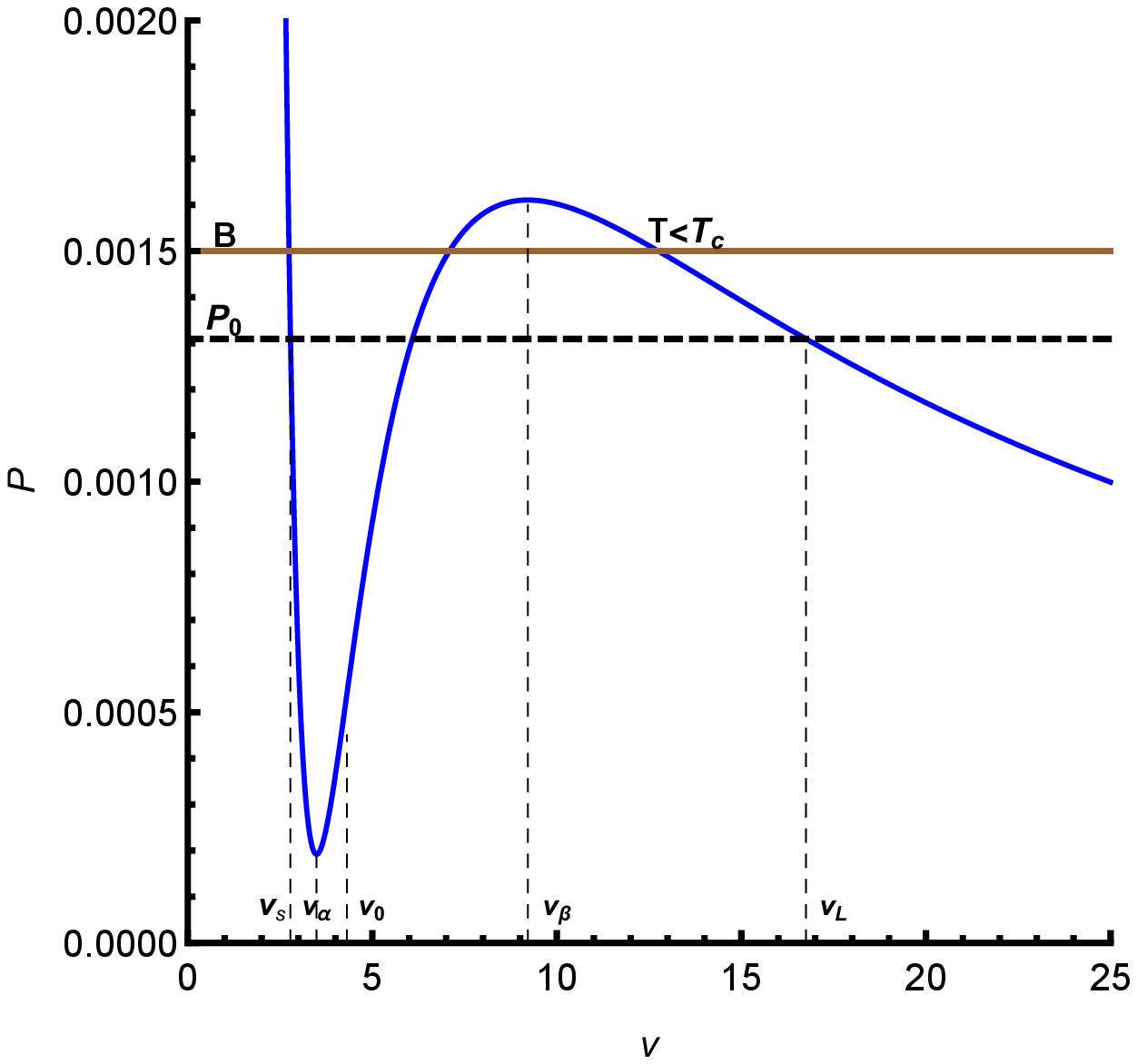}\;\;\includegraphics[width=5cm]{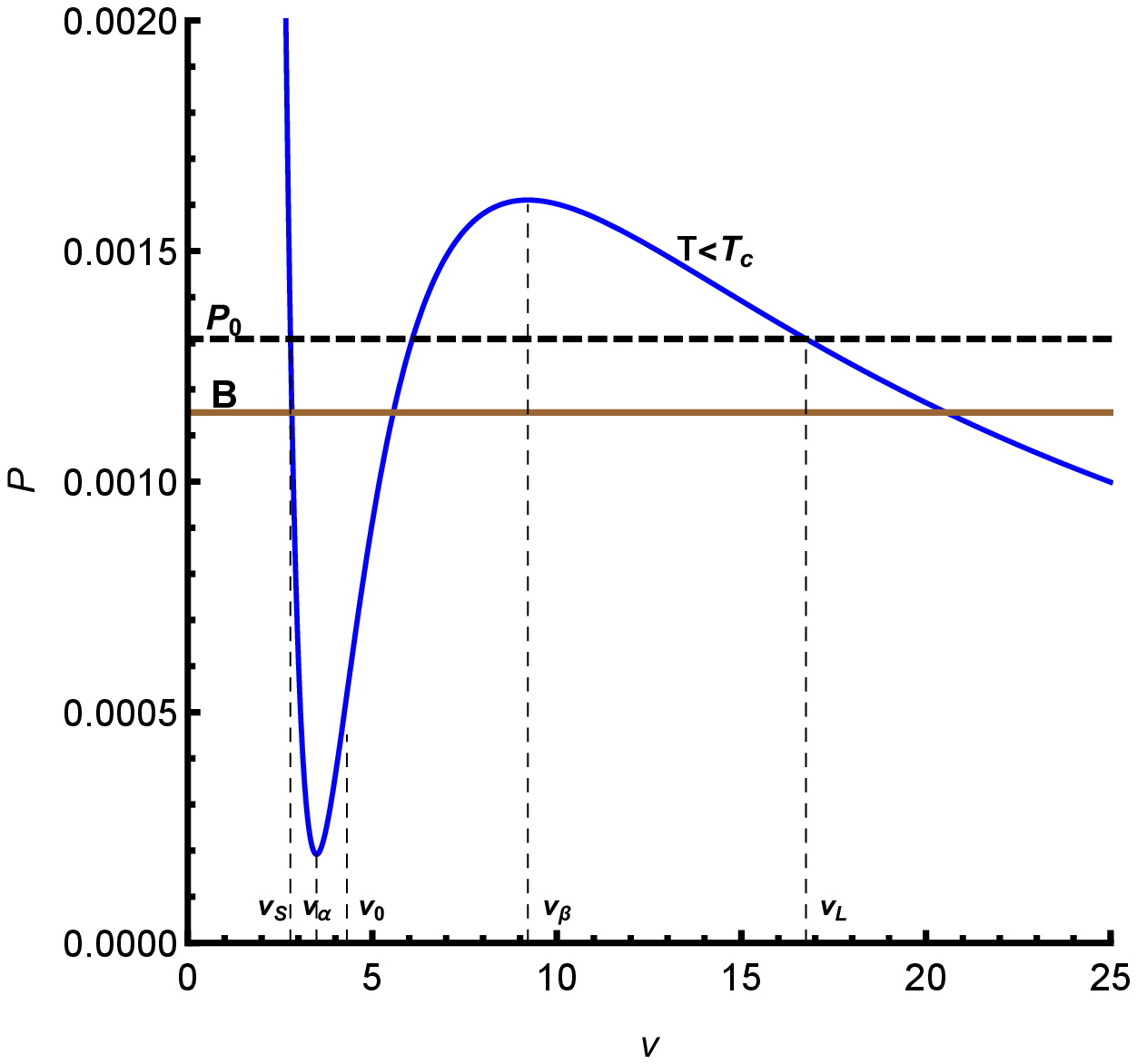}\;\;
\includegraphics[width=5cm]{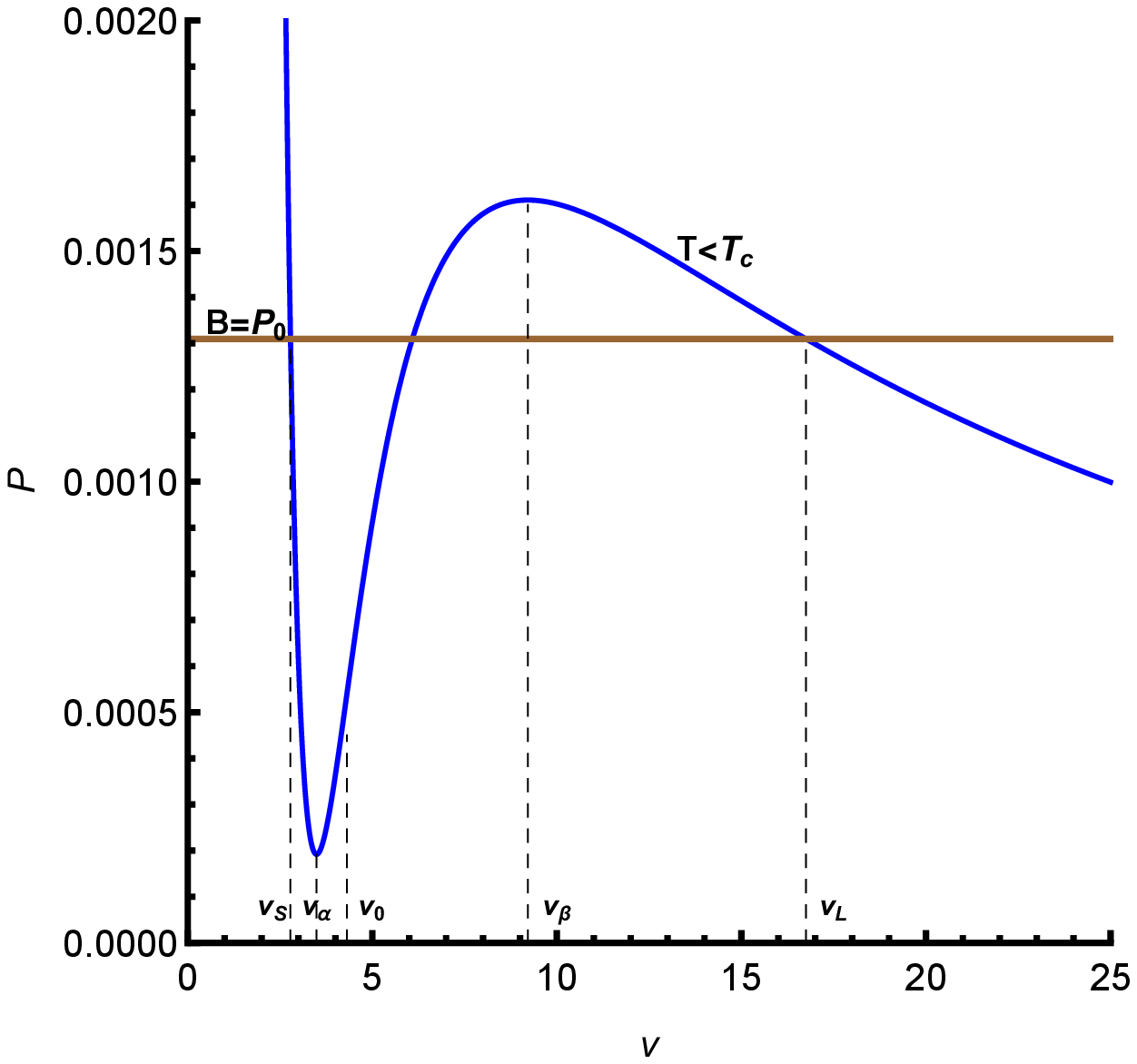}\\
\includegraphics[width=5cm]{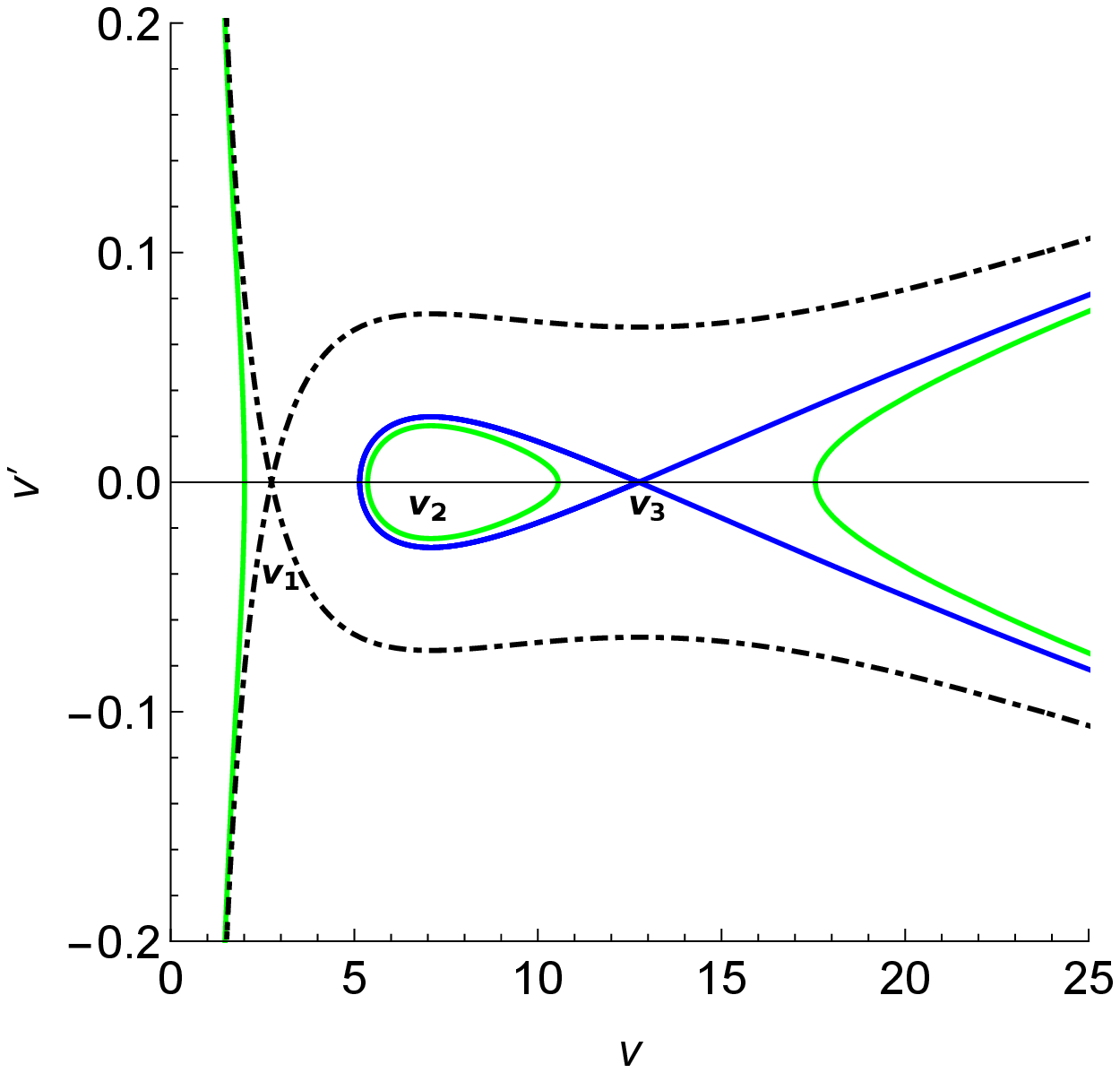}\;\;\includegraphics[width=5cm]{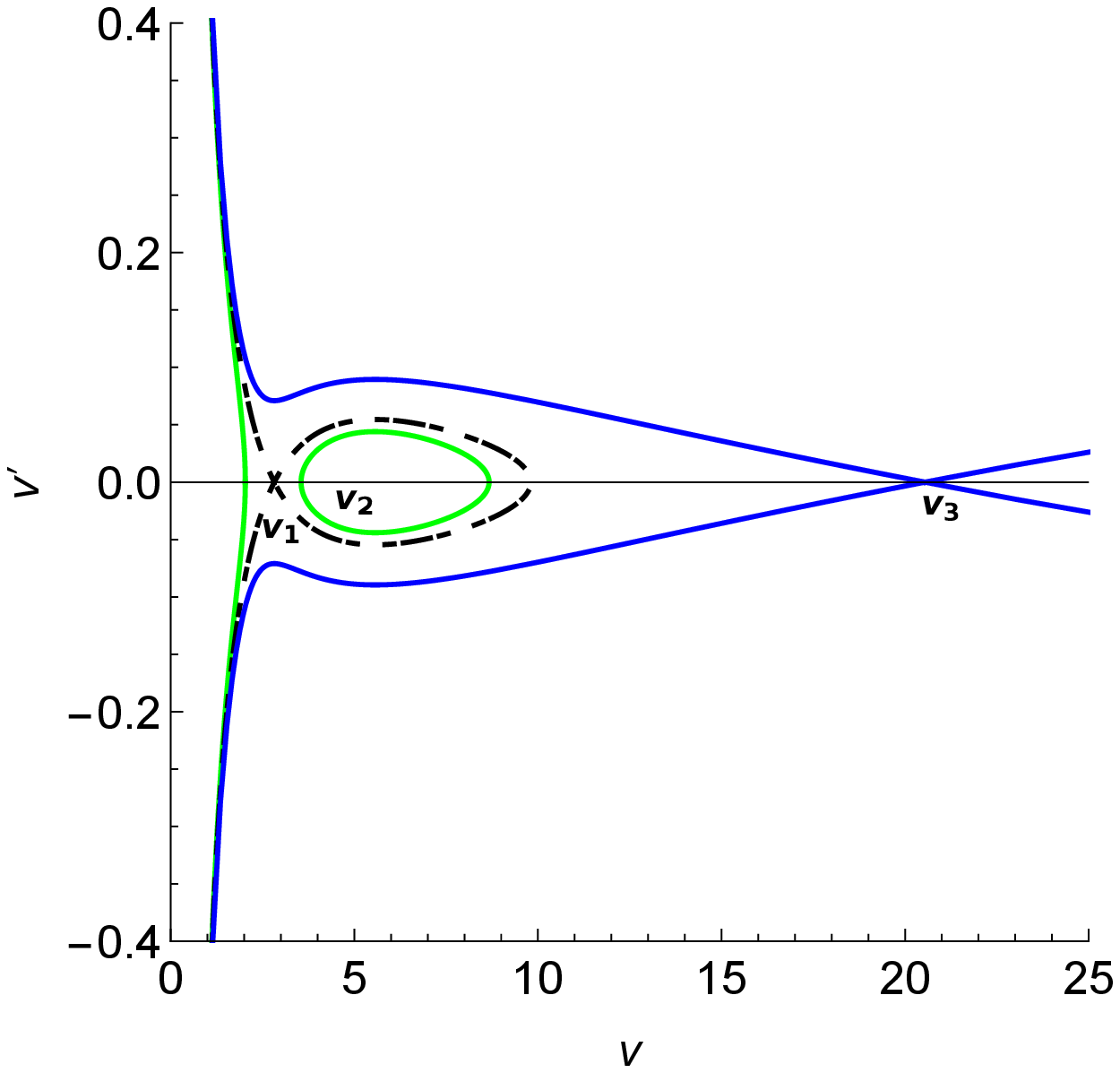}\;\;
\includegraphics[width=5cm]{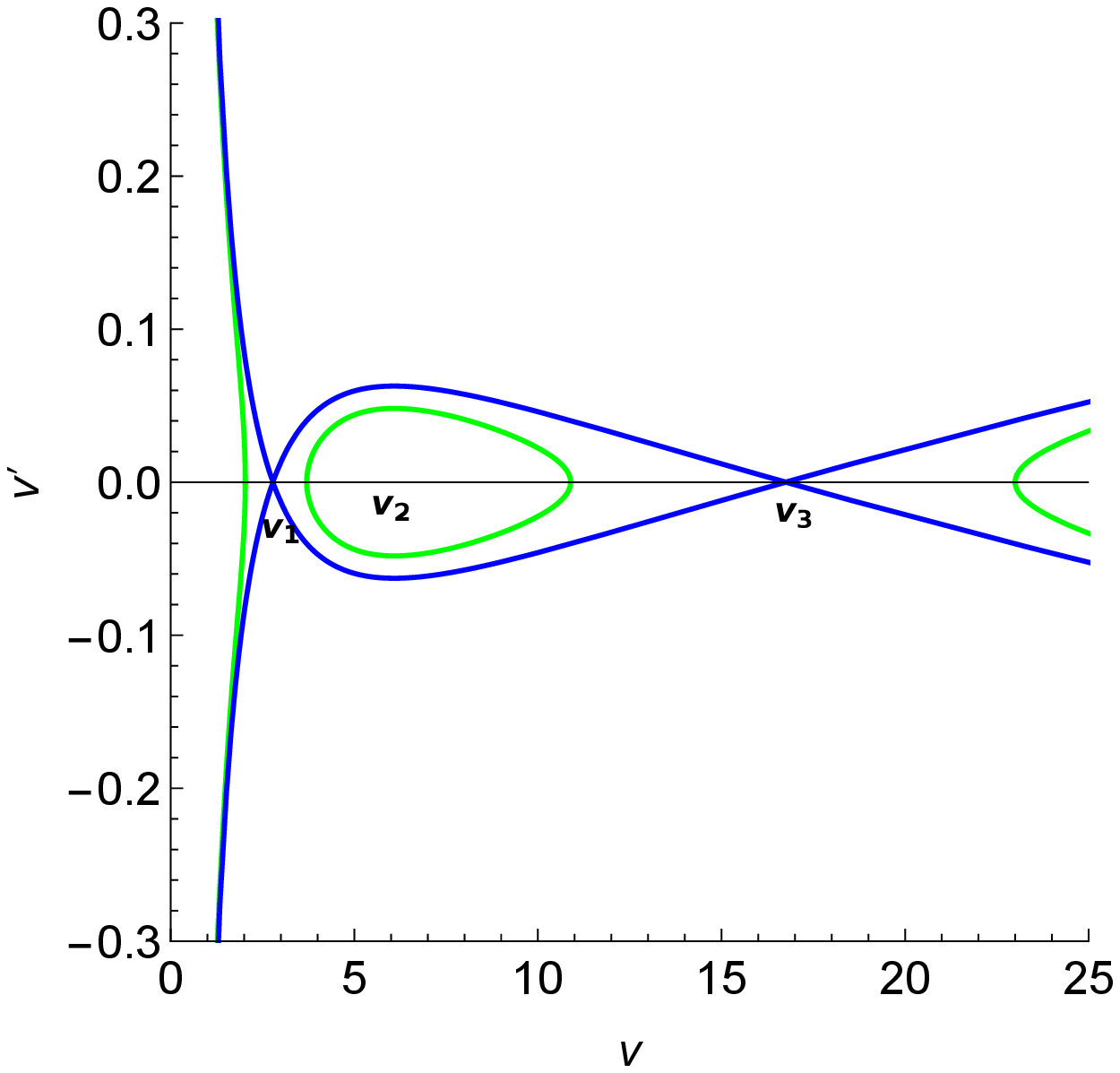}
\caption{ $P-v$ curve (the top row) and the corresponding $v-v'$ phase portrait (the bottom row). The left, the middle and the right panels correspond to the cases with $P_0<B<P(v_{\beta},T_0)$, $P(v_{\alpha},T_0)<B<P_0$, and $B=P_0$, respectively.  The blue lines in phase portraits  denote the homoclinic or heteroclinic orbits connecting saddle points. Here, we set $\alpha=0.01$, $b=1$, $q=1$, $T_0 = 0.0313$,
 and $P_0 =0.0013$.}\label{FIG5}
\end{center}
\end{figure}
For any fixed temperature $T_0<T_c$, the non-linear systems in Eq. (\ref{MainEq2}) have three fixed points, which are located at $v=v_1$, $v_2$ and $v_3$, respectively. Comparing the value of the phase transition pressure $P_0$ with the ambient pressure $B$, we can find three different types of phase structures in the  phase plane $v-v'$ as illustrated in Fig.(\ref{FIG5}).
As the ambient pressure $B$ is in the range $P_0<B<P(v_{\beta},T_0)$, one can find that there is a homoclinic orbit connecting the saddle point $v=v_3$ to itself. Similarly, as $P(v_{\alpha},T_0)<B<P_0$,  there exist also a homoclinic orbit connecting $v_1$ to itself. However, $B=P_0$, there is a heteroclinic orbit connecting $v_1$ with $v_3$. Three different types of phase structures in the phase plane for the charged dilaton-AdS black hole are similar to those for the RN-AdS \cite{BIAdS7}, Gauss-Bonnet AdS black holes \cite{BIAdS11} and Born-Infeld-AdS \cite{BIAdS}, which could be regarded as a common feature of such kind of static AdS black holes. Supposing a spatially periodic thermal perturbation has a form \cite{BIAdS5,BIAdS6,BIAdS7,BIAdS11,BIAdS}
\begin{eqnarray}
T=T_0 +\epsilon \cos{p x},
\end{eqnarray}
one can find that the dynamical equation (\ref{MainEq2}) becomes
\begin{eqnarray}
Av''+P(v,T_0)+ \frac{\epsilon \cos{px}}{v}=B.\label{MainEq3}
\end{eqnarray}
This second-order differential equation can be rewritten as a pair of first-order differential equations
\begin{eqnarray}\label{xut2}
v'&=&u, \nonumber \\
u'&=&\frac{1}{A}[B-P(v,T_0)]-\frac{\epsilon \cos{px}}{A v}.
\end{eqnarray}
For the dynamical equation (\ref{MainEq3}), the general solutions describing the homoclinic or heteroclinic orbit can
 be expressed as
\begin{eqnarray}
	z=
	\left [
	\begin{array}{c}
		v_0(x-x_0)\\
		u_0(x-x_0)
	\end{array}
	\right ]. \label{fFunction2}
\end{eqnarray}
\begin{figure}[ht]
\begin{center}
\includegraphics[width=5cm]{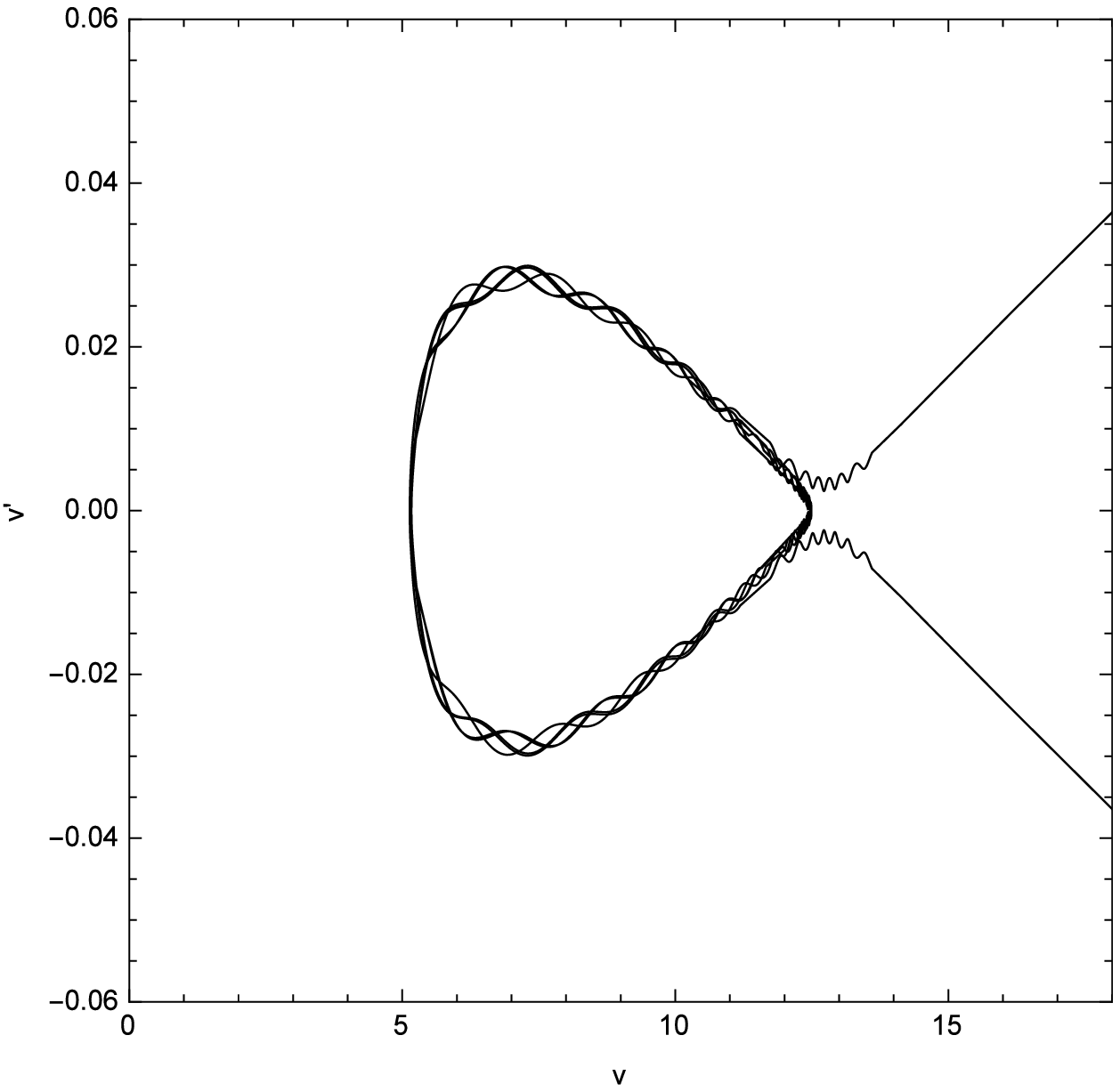}\;\;\includegraphics[width=5cm]{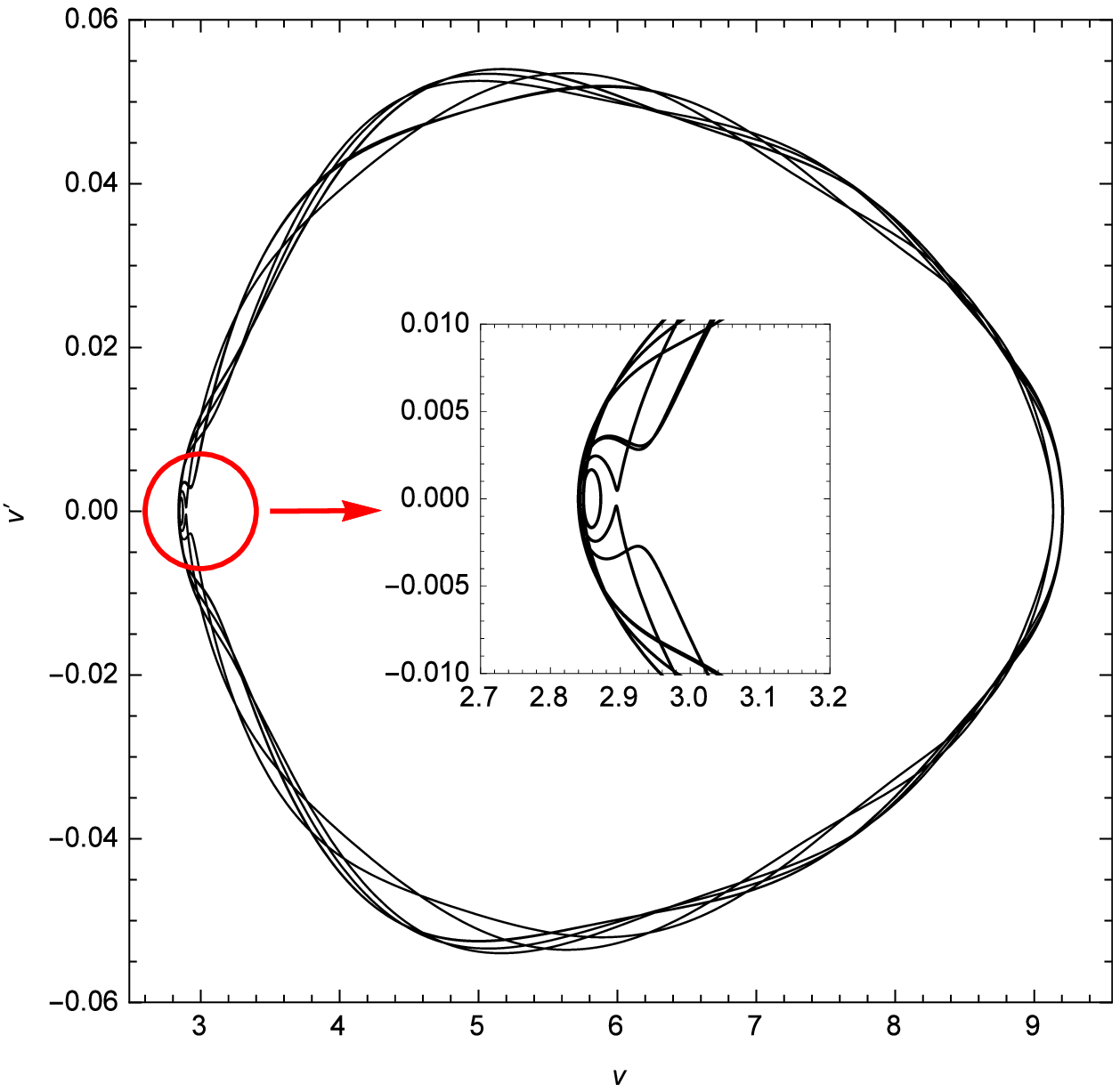}\;\;
\includegraphics[width=5cm]{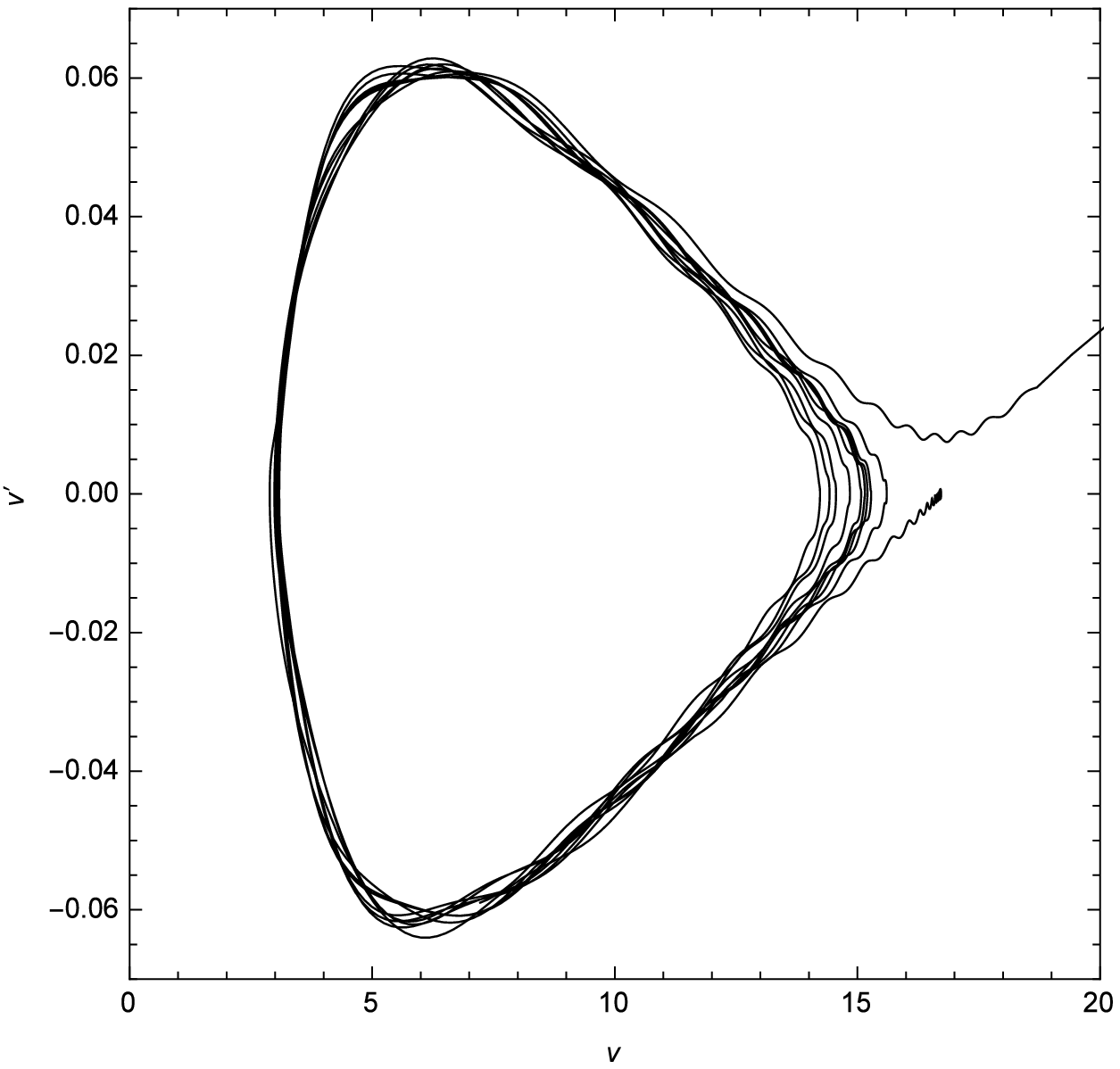}\\
\includegraphics[width=5cm]{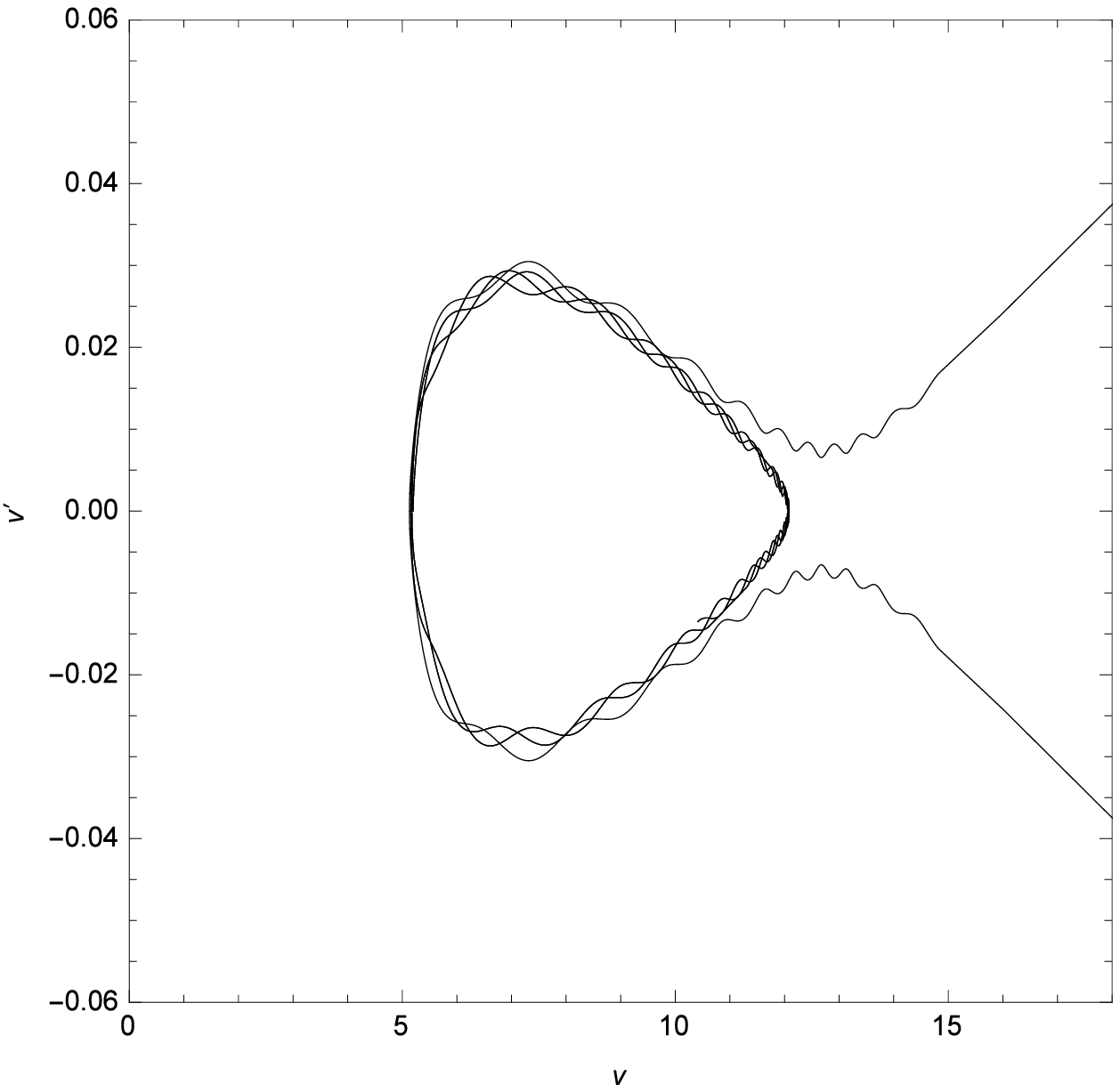}\;\;\includegraphics[width=5cm]{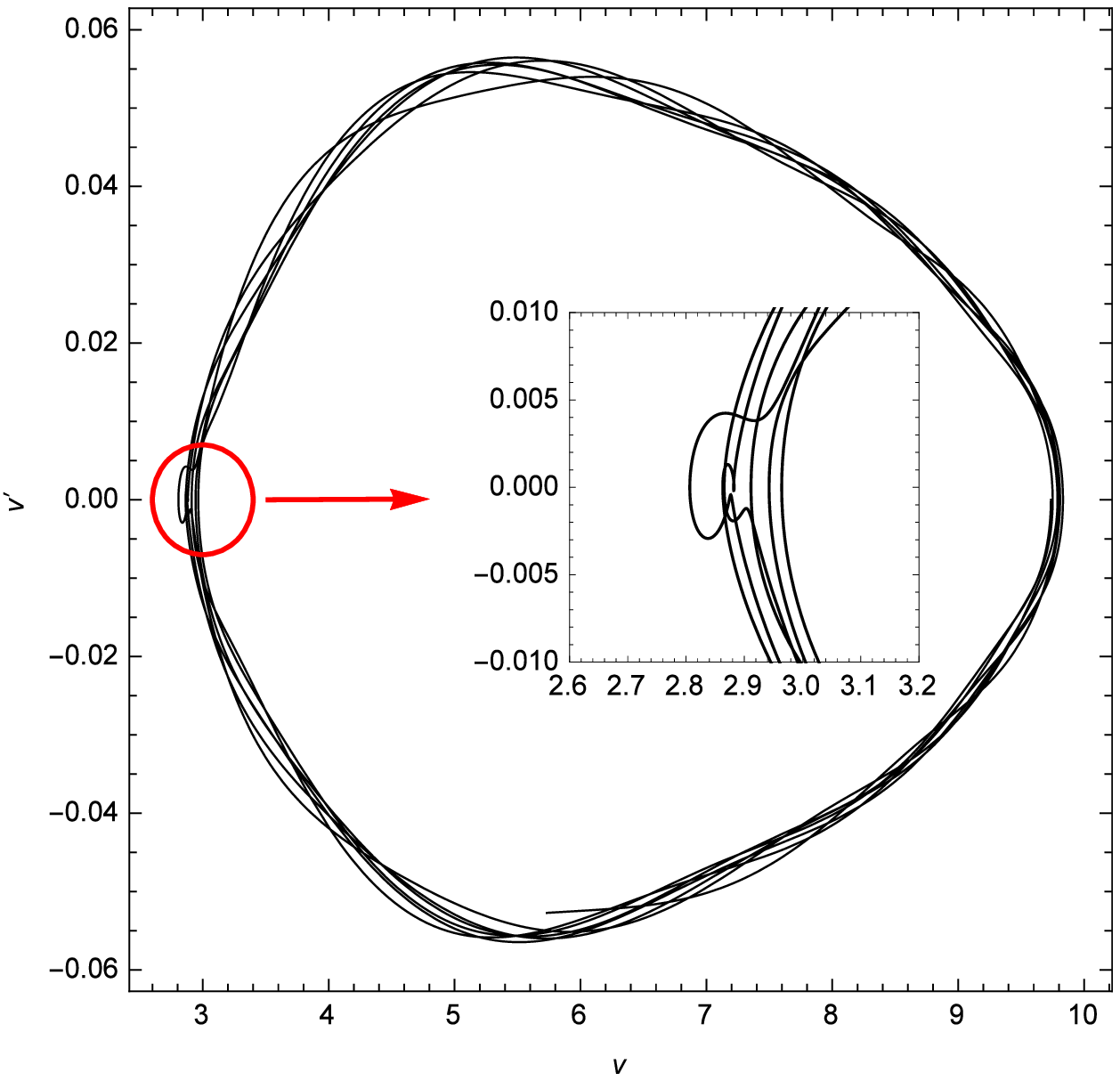}\;\;
\includegraphics[width=5cm]{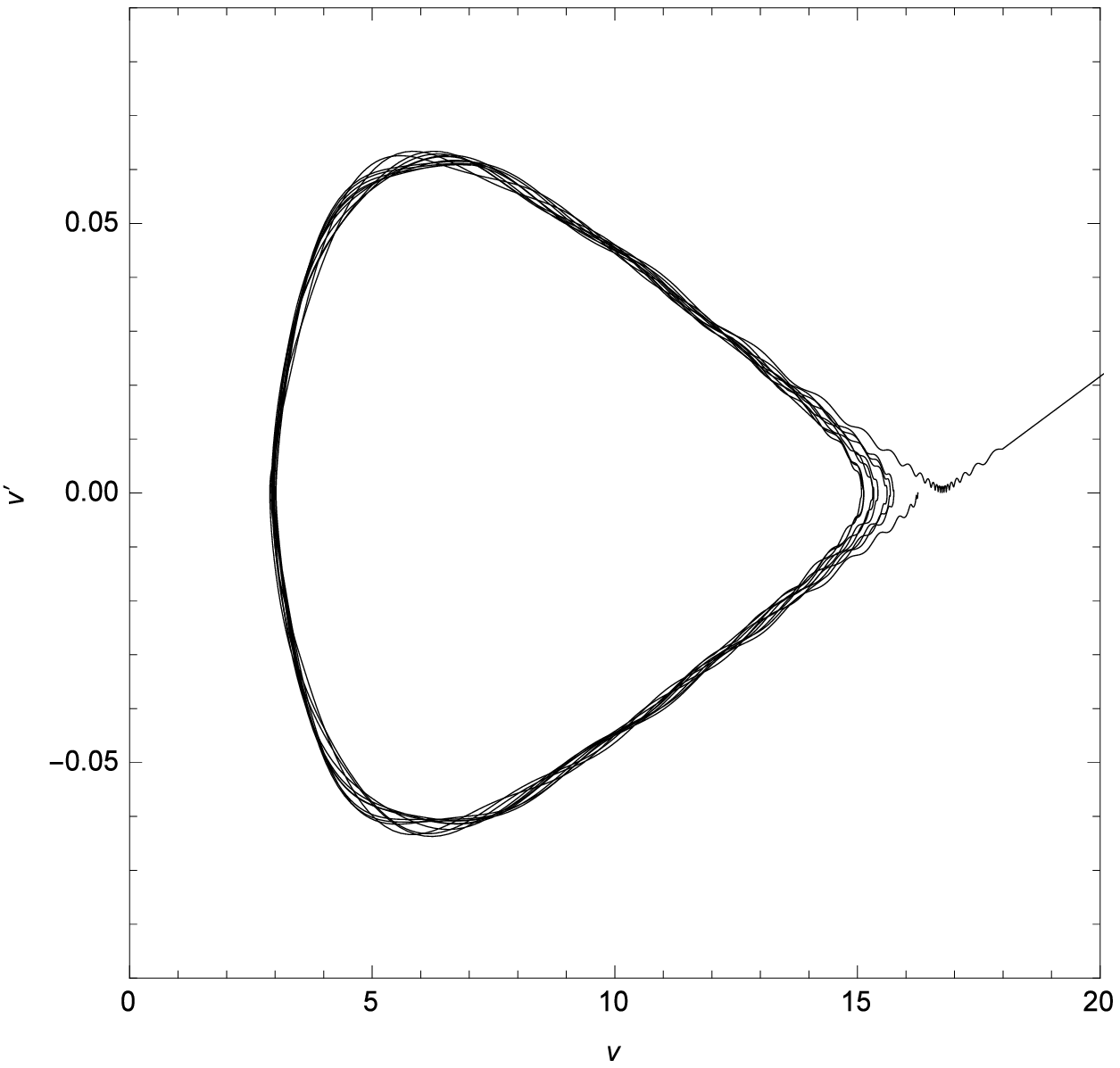}\\
\includegraphics[width=5cm]{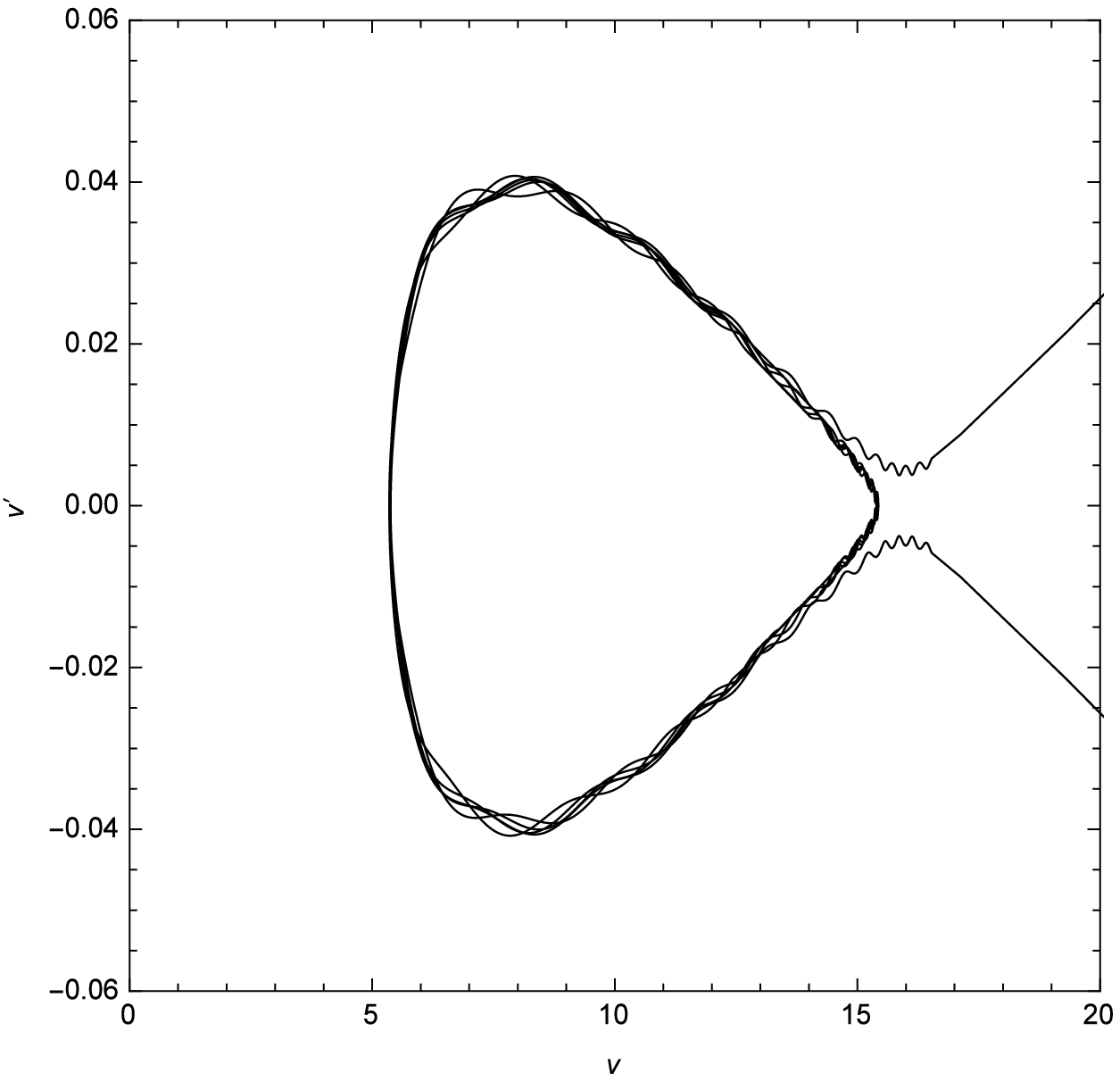}\;\;\includegraphics[width=5cm]{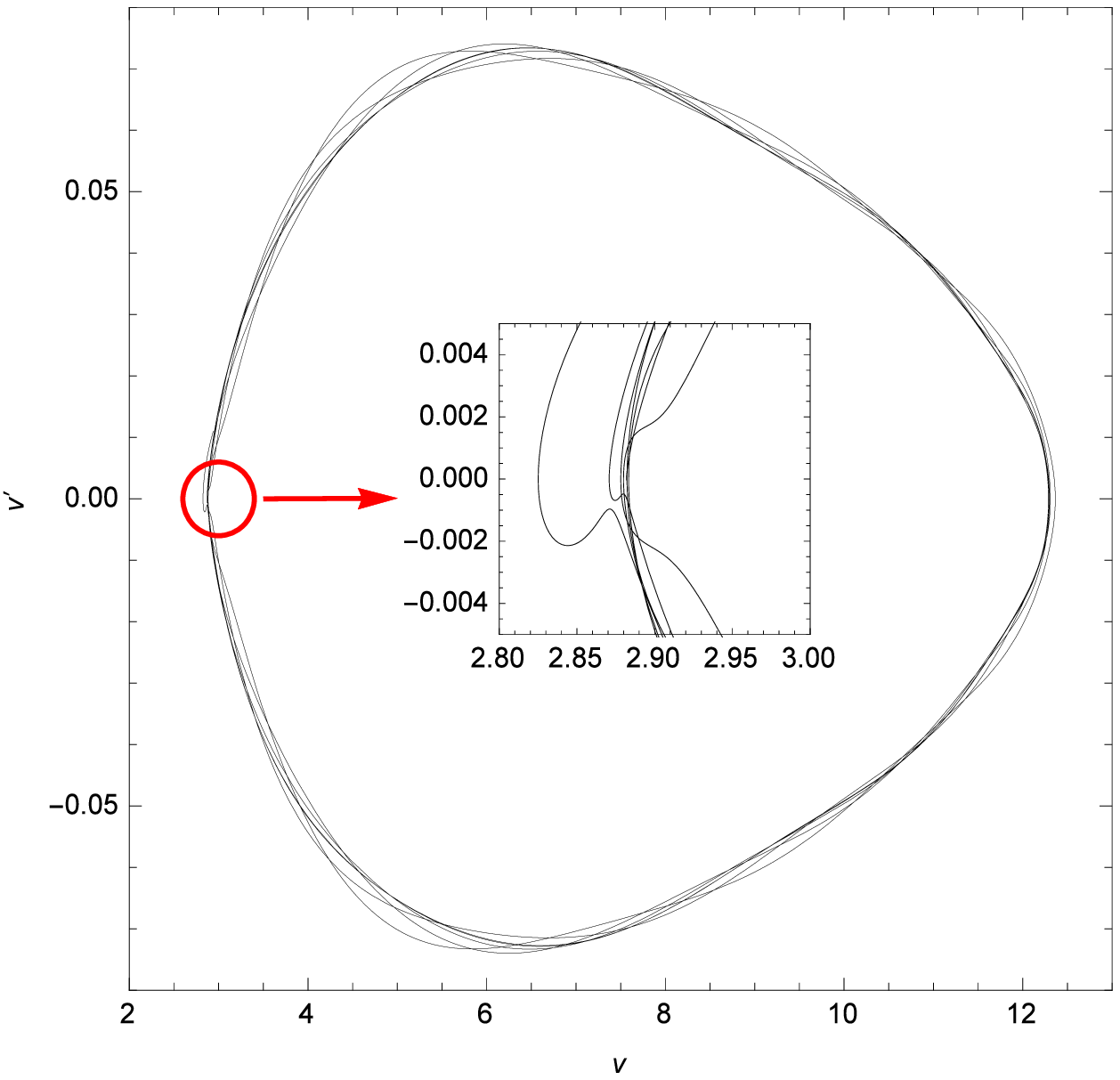}\;\;
\includegraphics[width=5cm]{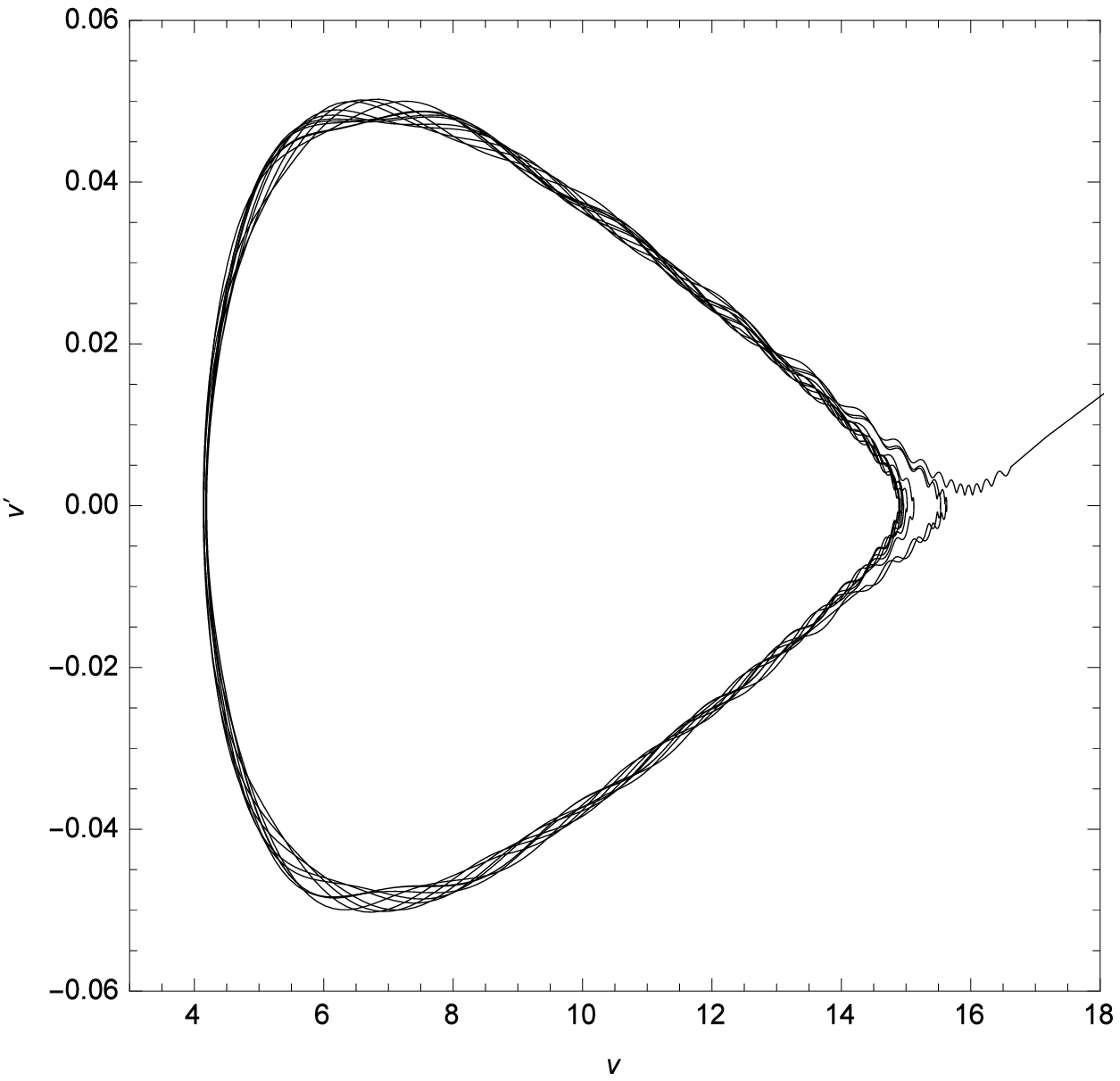}
\caption{ Phase portraits of the perturbed equation in $v-v'$  plane. The left, the middle and the right panels in every row correspond to the cases with $P_0<B<P(v_{\beta},T_0)$, $P(v_{\alpha},T_0)<B<P_0$ and $B=P_0$, respectively. The top, the middle and the bottom rows correspond to the cases with $\alpha=0$, $0.01$ and $0.1$,respectively. The initial states are the fixed points of the homoclinic orbit or heteroclinic orbit. Here, we set $b=1$, $q=1$, $T_0 = 0.0313$,
 and $P_0 =0.0013$. }\label{FIG8}
\end{center}
\end{figure}
and the corresponding functions $f(z(x-x_0))$ and $g(z(x-x_0),x)$ in Melnikov function becomes
\begin{eqnarray}
	&&f(z(x-x_0))=
	\left [
	\begin{array}{c}
		u_0(x-x_0) \\
		\frac{B}{A}-\frac{P(v_0(x-x_0),T_0)}{A}
	\end{array}
	\right ], \label{hFunction2}\\
 &&g(z(x-x_0),x)=
	\left [
	\begin{array}{c}
		0 \\
		-\frac{\cos{p(x-x_0)}}{A v_0(x-x_0)}
	\end{array}
	\right ].\label{gFunction2}
\end{eqnarray}
Thus, the Melnikov function in this cases can be simplified as
\begin{eqnarray}
M(x_0)=\int^{\infty}_{-\infty} -\frac{v'_0(x-x_0)\cos{p(x-x_0)}}{A v_0(x-x_0)}dx\equiv-\mathfrak{L}\cos{px_0}+N\sin{px_0},
\end{eqnarray}
with
\begin{eqnarray}
\mathfrak{L}=\int^{\infty}_{-\infty}\frac{v'_0(x-x_0)\cos{p x}}{v_0(x-x_0)}dx,\quad
N=\int^{\infty}_{-\infty}\frac{v'_0(x-x_0)\sin{p x}}{v_0(x-x_0)}dx.
\end{eqnarray}
Therefore, $M(x_0)$ always possesses simple zeros for the arbitrary values of $\mathfrak{L}$ and $N$. This means that there always exist chaos for thermodynamic system of the charged dilaton-AdS black hole suffered from a spatially periodic thermal perturbation. This is the same as that obtained in other AdS black holes \cite{BIAdS5,BIAdS7,BIAdS,BIAdS11}, which may be  a common feature of such kind of thermodynamic systems. In Fig. \ref{FIG8}, we chose the homoclinic orbit or heteroclinic orbit as the initial configures and plot the solutions of the perturbed equation (\ref{MainEq3}) in the $v-v'$ plane.  It is shown that there is spatial chaos under perturbations irrespective of black hole parameters and the perturbation strength.

\section{DISCUSSION AND SUMMARY}

We have studied thermal chaotic behavior in the extended phase space of a charged dilaton-AdS black hole by Melnikov method and present the effect of dilaton parameter on the thermal chaos. Under the temporal perturbation in the spinodal region, we find that the chaos occurs only if the perturbation amplitude $\delta$ is larger than a critical value $\delta_c$, which is similar to those in the RN-AdS, Born-Infeld-AdS  and Gauss-Bonnet AdS black holes. The dependence of $\delta_c$ on black hole parameters show that $\delta_c$ decreases with the charge $q$ and the parameter $b$, but increases with the dilaton parameter $\alpha$. This means that the presence of dilaton parameter makes the onset of chaos more difficult, which differs from those arising from the parameters $q$ and $b$.
For the spatially periodic thermal perturbation, we find that there always exists chaos for thermodynamic system of the charged dilaton-AdS black hole.
Comparing it with those of the RN-AdS, Born-Infeld-AdS  and Gauss-Bonnet AdS black holes, it is easy to obtain that for the temporal perturbation the thermal chaos
occurs only if the perturbation amplitude is larger than certain a critical value. However, for the spatially perturbation, the chaos always exists  irrespective of perturbation amplitude.
These behavior can be regarded as a common feature of such kind of static AdS black holes.

$\mathfrak{R}$

\section{\bf ACKNOWLEDGMENTS}
This work was partially supported by the National Natural Science Foundation of China under Grant No. 11875026, the Scientific Research
Fund of Hunan Provincial Education Department Grant
No. 17A124. J. Jing's work was partially supported by
the National Natural Science Foundation of China under
Grant No. 11875025.


\vspace*{0.2cm}
\begin{thebibliography}{99}

\baselineskip=0.6 cm \baselineskip=0.6 cm

\bibitem{Sprott} J. Sprott, \textit{Chaos and Time-Series Analysis}, Oxford University Press, 2003.
\bibitem{Ott} E. Ott, \textit{Chaos in Dynamical Systems}, Cambridge University Press, Second Edition 2002.
\bibitem{Brown1}R. Brown and L. Chua,  Int. J.
Bifurcation and Chaos {\bf6},  219 (1996); Int. J. Bifurcation and Chaos {\bf8}, 1 (1998).

\bibitem{Cornish} N.  Cornish, C. Dettmann and N. Frankel, Phys. Rev. D {\bf50} (1994) R618-621,[arXiv:gr-qc/9402027].

\bibitem{Hanan}W. Hanan and E. Radu, Mod. Phys. Lett. {\bf A} 22 (2007) 399-406, [gr-qc/0610119].

\bibitem{Contopoulos} J. Gair, C. Li, and I. Mandel, Phys. Rev. D {\bf77}, 024035 (2008).
\bibitem{Contopoulos1}G. Contopoulos, G. Gerakopoulos and T. Apostolatos, Int. J. Bifurc. Chaos {\bf21},  2261-2277  (2011); 
    
\bibitem{Contopoulos102}  G. Gerakopoulos, G. Contopoulos and T. Apostolatos,  Springer Proc.Phys. {\bf157} 129-136 (2014) arXiv:1408.4697.
\bibitem{Contopoulos2}F. Dubeibe, L. Pachon and J. Sanabria-Gomez, Phys. Rev. D {\bf 75}, 023008 (2007).
\bibitem{Contopoulos3}E. Gueron and P. Letelier, Phys. Rev. E {\bf66}, 046611 (2002).


\bibitem{Bombelli}L. Bombelli and E. Calzetta, Class. Quant. Grav. {\bf9}, 2573 (1992).
\bibitem{Bombelli1}J. Aguirregabiria, Phys. Lett. A {\bf224}, 234 (1997).
\bibitem{Bombelli2} Y. Sota, S. Suzuki and K. Maeda,  Class. Quant. Grav. {\bf13}, 1241 (1996).
\bibitem{Bombelli3}V. Witzany, O. Semerak and P. Sukova, Mon. Not. Roy. Astron. Soc. {\bf451} (2): 1770-1794 (2015).
\bibitem{Karas}V. Karas and D. Vokrouhlicky, Gen. Relativ. Gravit. {\bf24},729 (1992).


\bibitem{sbch} S. Chen, M. Wang and J. Jing, J. High Energy Phys.{\bf09}, 082 (2016).

\bibitem{Frolov}A.  Frolov and A.  Larsen,  Class. Quant. Grav. {\bf16 }, 3717-3724 (1999).
\bibitem{Zayas} L. Zayas and C. Terrero-Escalante, J. High Energy Phys. {\bf09}, 094 (2010).
\bibitem{MDZ}D. Ma, J. Wu and J.  Zhang, Phys. Rev. D {\bf89}, 086011 (2014).

\bibitem{BIAdS5} M. Slemrod and J. Marsden, Adv. Applied Math. {\bf6}, 135 (1985).

\bibitem{BIAdS6} V. Melnikov,  Trans. Mosc. Math. Soc. {\bf12}, 3 (1963).

\bibitem{R12} D. Kubiznak, R. Mann, J. High Energy Phys. {\bf07}, 033 (2012).

\bibitem{R120}S. Gunasekaran,  D. Kubiznak and R. Mann, J. High Energy Phys. {\bf11}, 110 (2012).
\bibitem{R121}R. Banerjee  and D. Roychowdhury, Phys. Rev. D {\bf85}
104043, (2012);  Phys. Rev. D {\bf85}, 044040 (2012).
\bibitem{R122} S. Wei and Y. Liu,  Phys. Rev. D {\bf87}, 044014 (2013).
\bibitem{R123} S. Hendi and M. Vahidinia, Phys. Rev. D {\bf88}, 084045 (2013)

\bibitem{BIAdS7} M. Chabab, H. Moumni, S. Iraoui, K. Masmar and S. Zhizeh, Phys. Lett. B {\bf781}, 316 (2018) [arXiv:1804.03960 [hep-th]].
\bibitem{BIAdS11} S. Mahish and B. Chandrasekhar,  Phys. Rev. D {\bf99}, 106012 (2019) [arXiv:1902.08932 [hep-th]].
\bibitem{BIAdS} Y. Chen, H. Li, S. Zhang, Gen. Rel. Grav. {\bf51}, 134 (2019) [arXiv:1907.08734 [hep-th]].

\bibitem{R3ad}  A. Sheykhi, Phys. Rev. D {\bf76}, 124025 (2007).

\bibitem{R1} A. Sheykhi, Phys. Lett. B {\bf662}, 7 (2008).

\bibitem{R2} A. Dehyadegari, A. Sheykhi and A. Montakhab, Phys. Rev. D {\bf96}, 084012 (2017).

 \bibitem{R3} M. Dehghani, S. Kamrani and A. Sheykhi, Phys. Rev. D {\bf90}, 104020 (2014).

 \bibitem{BIAdS10} B. Felderhof, Physica. D. {\bf48},  541 (1970).
\bibitem{BIAdS12} P. Holmes,  Phil. Trans. Roy. Soc. A {\bf292}, 419 (1979).
 \bibitem{BIAdS8} P. Holmes, Poincare,  Phys. Rep. {\bf193}, 137 (1990).

\bibitem{BIAdS9} V. Aslanov, \textit{Rigid body dynamics for space applications}, Butterworh-Heinemann Press (2017).
\bibitem{homo chaos} G. Cicogna and L. Fronzoni,  Phys. Rev. E {\bf47},4585. [arXiv:chao-dyn/9304006].

\end{thebibliography}
\end{document}